\documentclass[preprint,pr]{revtex4}
\usepackage{amsfonts}
\usepackage{amssymb}
\usepackage{color}  
\usepackage{graphicx}
\usepackage{epstopdf}

\newcommand{\Br}{{\bf r}}

\begin{document}

\title{Single-Scattering Optical Tomography: Simultaneous
Reconstruction of Scattering and Absorption}

\author{ Lucia Florescu$^{1}$, John C. Schotland$^{1}$, and Vadim A.
  Markel$^{2}$ }

\affiliation{$^{1}$Department of Bioengineering, University of Pennsylvania,
  Philadelphia, PA 19104, \\ 
$^{2}$Department of Radiology, University of Pennsylvania,
  Philadelphia, PA 19104}

\begin{abstract}
We demonstrate that simultaneous reconstruction of 
scattering and absorption of  a mesoscopic system using
angularly-resolved measurements of scattered light intensity is
possible. Image reconstruction is realized based on the algebraic
inversion of a generalized Radon transform relating the scattering
and absorption coefficients of the medium to the measured light
intensity and derived using the single-scattering approximation to
the radiative transport equation.

\end{abstract}
\date{\today} 

\maketitle

\section{Introduction}
\label{sec:intro}
There is considerable interest in the development of techniques for
 three-dimensional optical imaging of biological systems systems. In
 this context, of particular importance is the imaging of mesoscopic
 systems, where the photon transport mean free path is of the same
 order as the system size \cite{vanrossum_1999}. For such systems none
 of the available optical ballistic imaging modalities
 \cite{Wilson}-\cite{Ralston_2007} or the diffuse optical tomography
 \cite{Arridge_1999} can be used. On the mesoscopic length scale,
 applications to biological systems include imagining of engineered
 tissues, semitransparent organisms, or superficial tissues.  This
 article is the second in a series devoted to the problem of optical
 imaging in the mesoscopic scattering regime. In Ref.~\cite{SSOT1}, we
 have proposed a novel imaging technique that uses angularly-selective
 measurements of scattered light intensity to reconstruct the
 attenuation coefficient of an inhomogeneous medium, assuming that the
 single-light scattering is dominant. The advantages of this
 Single-Scattering Optical Tomography (SSOT) technique include the
 linearity, well-posedness, and two-dimensional character of the
 associated inverse problem, as well as the possibility to perform
 image reconstruction based on single-projection measurements. Here we
 generalize the SSOT technique to simultaneously reconstruct the
 internal scattering and absorption properties of the medium.

We begin by presenting a brief review of the SSOT formalism
introduced in Ref.~\cite{SSOT1}. We assume that the light transport in
an inhomogeneous medium is described by the time-independent
radiative transport equation (RTE)  for the specific intensity
$I({\bf r},\hat{\bf s})$ of light at the position ${\bf r}$ and
flowing in the direction $\hat{\bf s}$,
\begin{equation}
\label{RTE}
\left[ \hat{\bf s} \cdot \nabla + \mu_a({\bf r}) + \mu_s({\bf r}) \right] 
I({\bf r}, \hat{\bf s}) = 
\mu_s({\bf r}) \int A(\hat{\bf s},\hat{\bf s}^{\prime}) 
I({\bf r}, \hat{\bf s}^{\prime})d^2\hat{s}^{\prime} \ , \ \ {\bf r}\in
V \ .
\end{equation}
Here $\mu_a({\bf r})$ and $\mu_s({\bf r})$ are the absorption and
scattering coefficients, and $A(\hat{\bf s},\hat{\bf s}^{\prime})$ is
the scattering kernels normalized such that $\int A(\hat{\bf
s},\hat{\bf s}^{\prime}) d^2\hat{\bf s}^{\prime} = 1$ for all
$\hat{\bf s}$. The RTE (\ref{RTE}) is equivalent to the
integral equation
\begin{equation}
\label{RTE_int}
I({\bf r}, \hat{\bf s}) = I_b({\bf r}, \hat{\bf s}) + \int G_b({\bf
  r},\hat{\bf s}; {\bf r}^{\prime},\hat{\bf s}^{\prime}) \mu_s({\bf
  r}^{\prime}) A(\hat{\bf s}^{\prime},\hat{\bf s}^{\prime\prime})
I({\bf r}^{\prime}, \hat{\bf s}^{\prime\prime}) d^3r^{\prime} d^2
\hat{s}^{\prime}d^2\hat{s}^{\prime\prime} \ ,
\end{equation}
where $I_b({\bf r},\hat{\bf s})$ is the ballistic component of the
specific intensity, and the ballistic Green's function $G_b({\bf r},\hat{\bf s}; {\bf
  r}^{\prime},\hat{\bf s}^{\prime})$ is  expressed as
\begin{equation}
\label{G}
G_{b}({\bf r},\hat{\bf s}; {\bf r}^{\prime},\hat{\bf s}^{\prime}) = 
g({\bf r}, {\bf r}^{\prime})
\delta\left(\hat{\bf
    s}^{\prime} - \frac{\Br-\Br'}{|\Br-\Br'|}\right) \delta(\hat{\bf s} - \hat{\bf s}^{\prime}) \,
, 
\end{equation}
with
\begin{equation}
\label{g}
g({\bf r}, {\bf r}^{\prime}) = 
{1 \over {\vert {\bf r} - {\bf r}^{\prime}
\vert^2}} \exp\left[ - \int_0^{\vert {\bf r} - {\bf r}^{\prime}\vert}
\mu_t\left({\bf r}^{\prime} + \ell \frac{\Br-\Br'}{|\Br-\Br'|}\right)d\ell \right] \ 
\end{equation}
 the angularly-averaged ballistic Green's function. Here 
$\mu_t({\bf r}) = \mu_a({\bf r}) + \mu_s({\bf r})$ is the extinction
(attenuation) coefficient.

\begin{figure}
\includegraphics[ width=0.8\linewidth]{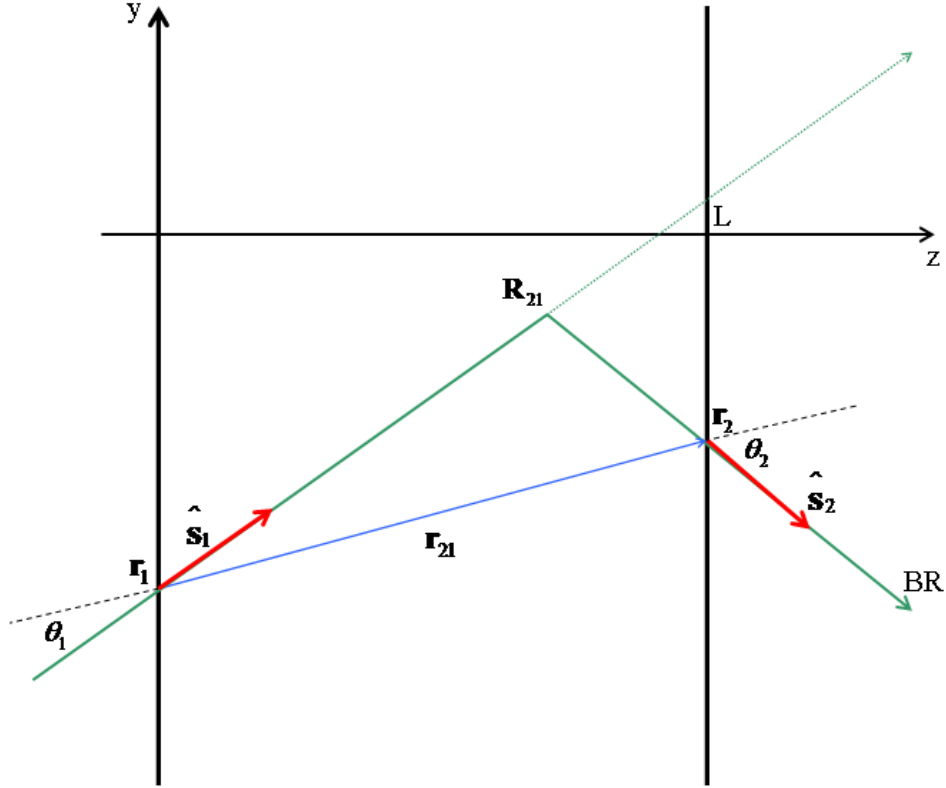}
\caption{\label{fig:sketch2} (Color online) Geometrical illustration
  of the quantities used in Eq.~(\ref{data_def}) and elsewhere. ``BR''
  denotes broken, single-scattered ray.}
\end{figure}

The light transport in a mesoscopic system is described by the
first-order scattering approximation to the RTE. This corresponds to
the assumption that light propagating in the inhomogeneous medium is
just single scattered, and consists in replacing $I({\bf r},\hat{\bf
s})$ by $I_b({\bf r},\hat{\bf s})$ in the right-hand side of
Eq.~(\ref{RTE_int}). Consider that the  medium is illuminated by
a light beam of intensity $I_0$ entering the slab at the point ${\bf
r}_1$ and in the direction $\hat{\bf s}_1$, and that an
angularly-selective detector registers the ray exiting the slab
through the opposite surface at the point ${\bf r}_2$ and in the
direction $\hat{\bf s}_2$ (as shown in Fig.~\ref{fig:sketch2}).  The
intensity measured in a such experiment is denoted by $I_s({\bf
r}_2,\hat{\bf s}_2; {\bf r}_1,\hat{\bf s}_1)$. Within the
single-scattering approximation, a relationship between the scattering
and absorption coefficients of the medium and the measured light
intensity is derived in the form

\begin{equation}
\label{SSOT_eq}
\int_{{\rm BR}({\bf r}_2,\hat{\bf s}_2;{\bf r}_1,\hat{\bf
    s}_1)}\mu_t[{\bf r}(\ell)] d\ell -\ln\left[ \frac{\mu_s({\bf
      R}_{21})}{\bar \mu_s} \right]
= \phi({\bf r}_2,\hat{\bf s}_2; {\bf r}_1,\hat{\bf s}_1) \ .
\end{equation}
Here the integral $\int_{\rm SSR}\mu_t({\bf r}(\ell)) d\ell$ of the
attenuation function is evaluated along the {\em broken ray} (BR)
(shown in Fig.~\ref{fig:sketch2}), corresponding to single-scattered
photons and uniquely defined by the source and detector positions and
orientations, $\ell$ is the linear coordinate on this ray, ${\bf
  R}_{21}$ is the ray turning point, and  $\bar \mu_s $ is  the average
(background) value of the scattering coefficient.  The data function $\phi({\bf r}_2,\hat{\bf s}_2; {\bf
  r}_1,\hat{\bf s}_1)$ is defined as
 \begin{equation}
\label{data_def}
\phi({\bf r}_2,\hat{\bf s}_2; {\bf r}_1,\hat{\bf s}_1) = 
-\ln\left[ \frac{r_{21} \sin\theta_1 \sin\theta_2 \int 
I_s({\bf r}_2,\hat{\bf s}_2; {\bf r}_1,\hat{\bf s}_1)
d\varphi_{\hat{\bf s}_2}}{I_0 \bar \mu_s   A(\hat{\bf s}_2, \hat{\bf s}_1)}
\right] \ ,
\end{equation}
where $r_{21}=\vert{\bf r}_2 - {\bf r}_1 \vert$, the angles $\theta_1$
and $\theta_2$ are defined by $\cos\theta_{1,2}=\hat{\bf
  r}_{21}\cdot\hat{\bf s}_{1,2}$, $\varphi_{\hat{\bf s}_2}$ is the
polar angle of $\hat{\bf s}_2 $, and the scattering kernel $A(\hat{\bf s},\hat{\bf
  s}^{\prime})$ is assumed position-independent and known.

Eq.~(\ref{SSOT_eq}) is applied to optical imaging in the following
manner. The selection of incidence and detection points and incidence
and detection directions defines a slice in which image reconstruction
is performed. In Fig.~\ref{fig:sketch2}, this slice coincides with the
$YZ$-plane of the laboratory frame.  Assuming that the $x$-coordinate
is fixed, the absorption and scattering coefficients can be regarded
as two-dimensional functions of variables $(y,z)$. On the other hand,
the data are four-dimensional, depending, in general, on two spatial
and two angular variables, corresponding to the source and detection
y-coordinate, and source and detection direction, respectively. This
enables the simultaneous reconstruction of scattering and absorption.
By utilizing multiple incident beams and detecting light exiting the
medium at different points, and by varying the incident and exit
angles, it is possible to collect enough data to reconstruct the
absorption and scattering coefficients in a given slice.
Three-dimensional reconstruction is then performed slice-by-slice.

\begin{figure}
\includegraphics[ width=0.45\linewidth]{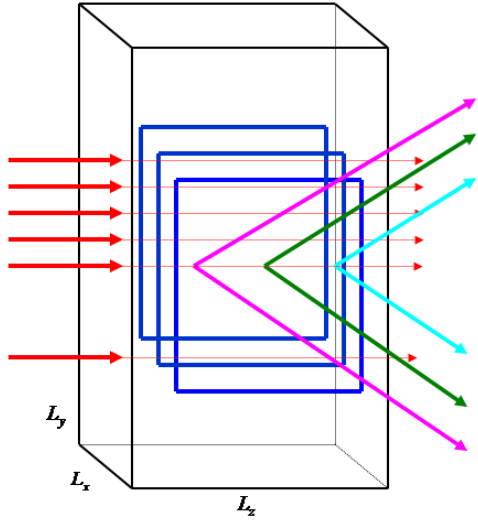}
\caption{(Color online) Schematic illustration of the proposed
  experiment geometry for simultaneous reconstruction of scattering
  and absorption.  Reconstruction is performed  in slices distributed along the $x$ direction. 
 The blue rectangles represent the areas in which reconstruction can
  be performed.}
\label{fig:sketch1}
\end{figure}

In SSOT, simultaneous reconstruction of scattering and absorption can
be in fact realized without scanning all parameter space. It is enough
to keep the incidence direction (defined by the incidence angle
$\beta_1$) fixed, to scan the incidence point $y_1$, and for each such
source realization to scan the detection point $y_2$, for each
detection position considering just one detection direction,
corresponding to the detection angle $\beta_2>\beta_1$, if $y_2>
y_1+L\tan\beta_1$, or to the angle $-\beta_2$, if $y_2<
y_1+L\tan\beta_1$, where $\beta_2$ is fixed and $L$ is the slab
thickness. The incidence and detection angles are the angles between
the $z$-axis of the laboratory frame and the unit vectors $\hat{\bf
s}_1$ and $\hat{\bf s}_2$, respectively.  This principle is
schematically illustrated in Fig.~\ref{fig:sketch1} for a rectangular
sample illuminated by a normally incident beam.  In the presence of
scattering, both ballistic and scattered rays are present. To avoid
the detection of the ballistic component of the transmitted light, the
angularly-selective source and detectors are not aligned with each
other.

For a light beam entering the sample at position $(y_1, z_1)$ and at
the incident angle $\beta_1$, consider pairs of detections defined by
the detection position and detection angle $(y_2, z_2)$ and
$\beta_2>\beta_1$, and $(y_{2}^{'}, z_2)$ and
$-\beta_2$, respectively, where $y_2> y_1+L\tan\beta_1$,
$y_{2}^{'}=y2-2(z_2-z_0)\tan\beta_2$, and
$z_0=(z_2\tan\beta_2-z_1\tan\beta_1)/(\tan\beta_2-\tan\beta_1)-(y_2-y_1)/(\tan\beta_2-\tan\beta_1)$.
This two measurements corresponds to rays single-scattered at the same
position ${\bf R}_{21}=(y_0,z_0)$ within the sample, where
$y_0=(z_2-z_0)\tan\beta_2$. Such rays are shown by the same color in
Fig.~\ref{fig:sketch1}. Writing Eq.~(\ref{SSOT_eq}) for the two
situations and then subtracting the corresponding equations, one
obtains the following equation:
\begin{equation}
\label{SIM_SSOT_eq1}
\int_{{\rm BR}(y_2, \beta_2; y_1, \beta_1)}\mu_t[y(\ell),z(\ell ] d\ell -\int_{{\rm BR}(y_{2}^{'}, -\beta_2; y_1, \beta_1)}\mu_t[y(\ell),z(\ell ] d\ell
= \phi(y_2, \beta_2; y_1, \beta_1)-\phi(y_{2}^{'}, -\beta_2; y_1, \beta_1).
\end{equation}
Here the integrals of the total attenuation coefficient are along the
two single-scattered rays described above, and $\phi(y_2, \beta_2;
y_1, \beta_1)$ and $\phi(y_{2}^{'}, -\beta_2; y_1, \beta_1)$ are the
data functions corresponding to the two detections.
Eq.~(\ref{SIM_SSOT_eq1}) can be solved with respect to $\mu_t(y,z)$.
Then  Eq.~(\ref{SSOT_eq}) written for one of
the detections can be solved for $\ln[\mu_s(y,z)/\langle
\mu_s\rangle]$.  From these two solutions, $\mu_a(y,z)$ and
$\mu_s(y,z)$ can be  determined.

\section{Image Reconstruction}
\label{sec:simulations}

In what follows, we illustrate simultaneous reconstruction of
scattering and absorption in SSOT based on the algebraic inversion of
Eq.~(\ref{SSOT_eq}). We note that a more sophisticated image
reconstruction algorithm based on a inverse formula is also possible
and will be presented elsewhere.

The forward data is obtained by solving Eq.~(\ref{RTE_int})
numerically, along the lines presented in detail in Ref.~\cite{SSOT1},
generalized here for the case of a homogeneously scattering medium.  For
isotropic scattering ($A(\hat{\bf s}, \hat{\bf s}) = 1/4\pi$),
Eq.~(\ref{RTE_int}) takes the form

\begin{equation}
\label{I_u}
I({\bf r},\hat{\bf s}) = I_b({\bf r},\hat{\bf s}) + \int G_b({\bf
  r},\hat{\bf s}; {\bf r}^{\prime},\hat{\bf s}^{\prime})
\frac{\mu_s({\bf r}^{\prime})}{4\pi}u({\bf r}^{\prime}) d^3r^{\prime}
d^2\hat{s}^{\prime}.
\end{equation}
Here $u({\bf r})\equiv \int I({\bf r},\hat{\bf s})
d^2\hat{s}$ is the density of electromagnetic energy satisfying the
following integral equation 

\begin{equation}
\label{u_u_b}
u({\bf r}) = u_b({\bf r}) + \int g_b({\bf r},{\bf r}^{\prime})
\frac{\mu_s({\bf r}^{\prime})}{4\pi}u({\bf r}^{\prime}) d^3r^{\prime}
\ , 
\end{equation}

\noindent
where $u_b({\bf r})\equiv \int I_b({\bf r},\hat{\bf s})d^2s$
is the ``ballistic density''.  The scattered component of the intensity,
$ I_s=I-I_b$, is computed by solving first Eq.~(\ref{u_u_b}) and then
substituting the numerical solution $u({\bf r})$ into (\ref{I_u}). We
emphasize that this numerical approach is non-perturbative and
includes all scattering orders, similarly to the experimental
situation when all scattered light is detected.

Equation~(\ref{u_u_b}) is discretized on a rectangular grid and
solved by methods of linear algebra. The energy density $u({\bf r})$
and the scattering and absorption coefficients are assumed constant
within each cubic cell. The corresponding
values $u_n=u({\bf r}_n)$, where ${\bf r}_n$ is the center of the
$n$-th cubic cell, obey the algebraic system of equations
\begin{equation}
\label{u_u_b_disc}
\left(1 - R_{\rm eq} \mu_s({\bf r}_n)\right) u_n - \frac{ h^3}{4\pi}\sum_{m\neq n}g_b({\bf
  r}_n, {\bf r}_m)\mu_s({\bf r}_m) u_m = u_b({\bf r}_n) \ .
\end{equation}
\noindent
Here $h$ is the disctretization step, $u_b({\bf r}_n) \equiv
h^{-3}\int_{V_n} u_b({\bf r}) d^3r$, and $R_{\rm eq}=(3/4\pi)^{1/3}h$
is the radius of a sphere of equivalent volume to that of a cell,
introduced to compute the diagonal matrix elements of the system
(\ref{u_u_b_disc}) \cite{SSOT1}.  The system of equations
(\ref{u_u_b_disc}) is solved by direct matrix inversion, and then the
specific intensity is calculated according with the discretized
version of (\ref{I_u}),
\begin{equation}
\label{I_u_d}
I({\bf r}_2, \hat{\bf s}_2) = \frac{h^3}{4\pi} 
\sum_{{\bf r}_2 - {\bf r}_n=\hat{\bf s}_2|{\bf r}_2 - {\bf r}_n|}
 g_b({\bf r}_2, {\bf r}_n)\mu_s({\bf r}_n) u_n \ ,
\end{equation}
where the summation is performed only over such cells that are
intersected by the ray exiting from the detection point ${\bf r}_2$ in
the direction $\hat{\bf s}_2$. The data function is calculated in
terms of the average of the specific intensity over the cell, ${\bar
I}({\bf r}_2, \hat{\bf s}_2)\equiv(1/h^3)\int d{\bf r}_2\, I({\bf
r}_2, \hat{\bf s}_2)$.  To model noise in the measured data, ${\bar
I}({\bf r}_2, \hat{\bf s}_2)$ was scaled and rounded off so that it
was represented by 16-bit unsigned integers, similar to the
measurement by digital ccd cameras. Then a statistically-independent
positively-defined random variable was added to each measurement
${\bar I}({\bf r}_2, \hat{\bf s}_2)$. The random variables were evenly
distributed in the interval $[0,n I_{\rm av}]$, where $n$ is the noise
level and $I_{\rm av}$ is the average measured intensity (a 16-bit
integer). The date function is calculated using the discretized
version of Eq.~(\ref{data_def}),
\begin{equation}
\label{data_discret}
\phi(y_2,\beta_2; y_1,\beta_1) = 
-\ln\left[\frac{4\pi}{h^3}\,\frac{{\bar I}({\bf r}_2, \hat{\bf s}_2)}{I_0\bar \mu_s }\right]. 
\end{equation}

Image reconstruction for the attenuation coefficient is obtained using
Eqs.~(\ref{SSOT_eq}) and (\ref{SIM_SSOT_eq1}), which are discretized on
the same grid as the one used for obtaining the forward solution,
except that in this case planar slices with fixed $x$-coordinates are
used. The discrete version of (\ref{SIM_SSOT_eq1}) is

\begin{equation}
\label{SSOT_2D_d}
\sum_n {\mathcal L}_{\nu n} \mu_{tn} = \phi_\nu \,,
\end{equation}
where the index $\nu=(y_1,\beta_1; y_2, \beta_2)$ corresponds to a
given realization of the source and detection pair, and ${\mathcal
L}_{\nu n}={\mathcal L}^{(1)}_{\nu n}-{\mathcal L}^{(2)}_{\nu n}$,
with the matrix element ${\mathcal L}^{(i)}_{\nu n}$ given by the
length of the intersection of the detected ray $i$ with the $n$-th
cubic cell.  $\phi_\nu= \phi_\nu^{(1)}-\phi_\nu^{(2)}$, with
$\phi_\nu^{(i)}$ the data function corresponding to the ray $i$.
Eq.~(\ref{SSOT_2D_d}) is solved for $\mu_{tn}$ by regularized SVD
pseudoinverse~\cite{natterer_book_01}, namely
\begin{equation}
\label{mu_SVD}
\vert \mu_t^+ \rangle = ({\mathcal L}^* {\mathcal L})^{-1} {\mathcal
  L}^* \vert \phi \rangle \ ,
\end{equation}

\noindent
where

\begin{equation}
\label{L_star_L_inv}
({\mathcal L}^* {\mathcal L})^{-1} = \sum_n \Theta(\sigma_n^2 -
\epsilon) \frac {\vert f_n \rangle \langle f_n \vert} {\sigma_n^2} \ .
\end{equation}

\noindent
Here $\Theta(x)$ is the step function, $\epsilon$ is a small
regularization parameter, and $\vert f_n \rangle$ and $\sigma_n$ are
the singular functions and singular values, respectively, of the
matrix ${\mathcal L}$, obtained  by solving the symmetric eigenproblem
${\mathcal L}^* {\mathcal L}\vert g_n \rangle = \sigma_n^2 \vert g_n
\rangle$.  Further, the scattering coefficient is determined from
Eq.~(\ref{SSOT_eq}), discretized as
\begin{equation}
\label{SSOT_2D_d}\sum_n {\mathcal L}^{(1)}_{\nu n} \mu_{tn}-\ln\left[ \frac{\mu_s({\bf
      R}_{21})}{\bar \mu_s} \right] = \phi^{(1)}_\nu(y_2,\beta_2; y_1,\beta_1). \,
\end{equation}
Finally, the absorption coefficient is obtained  as $\mu_{an}=\mu_{tn}-\mu_{sn}$.

\subsection{Numerical Results}
\label{results}

We considered a rectangular isotropically scattering sample of
dimensions $L_x=25h$, $L_y=122h$ and $L_z=40h$.  The background
scattering coefficient is chosen such that the optical depth
$\bar\mu_s L_z$ is $1.6$.  This corresponds to the mesoscopic
scattering regime in which the image reconstruction method of SSOT is
applicable. The background absorption coefficient is set equal to
$\bar\mu_a=0.1\bar\mu_s$. The target is a set of inclusions
concentrated in the layers $x=6h$, $x=13h$ and $x=20h$.

Image reconstruction is performed in slices $x=x_{\rm slice}={\rm
  const}$ separated by the distance $\Delta x=h$. The reconstruction
  area inside each slice is $44h\leq y \leq 77h$, $4h\leq z \leq 37h$,
  with the field of view $34h \times 34h$.  For each slice, the
  sources are normally incident ($\beta_1=0$) on the surface $z=0$ at
  the positions $x=x_{\rm slice}$, $y=y_s=nh$, $z=0$, with $n$
  integers.  The detectors are placed on the opposite side of the
  sample at positions $x=x_{\rm slice}$, $y=y_d=nh$, $z=L_z$ and
  measure the specific intensity exiting the surface $z=L_z$ at the
  angle of $\beta_2=\pi/4$, for $y_d>y_s$, and $-\pi/4$, for
  $y_d<y_s$, with respect to the $z$-axis.

\begin{figure}
\begin{center}
\includegraphics[height=10cm]{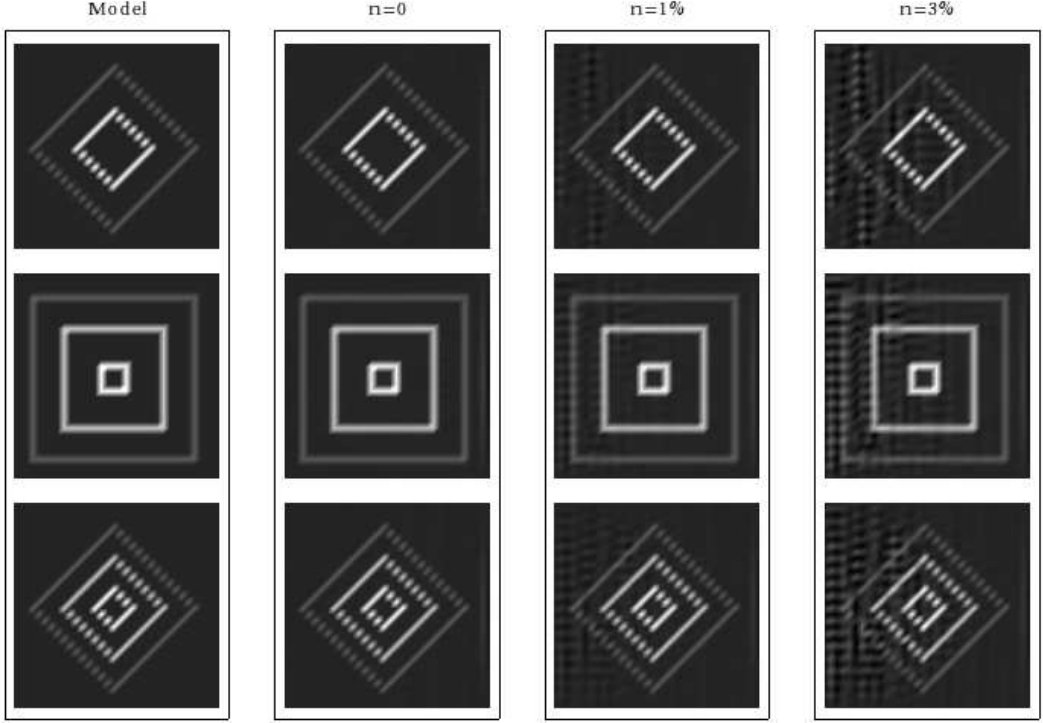}
\end{center}
\caption{\label{mu_t_h_0.04}(Color online) Image reconstruction for the total attenuation
  coefficient $\mu_t$ for a homogeneously scattering sample with
  $\bar\mu_{s}L_z=1.6$ and $\bar\mu_{a}=0.1\bar\mu_s$, and for various
  noise levels $n$. The rows show the slices $x=6h$, $13h$ and $20h$,
  where the absorbing inhomogeneities are placed. The contrast in
  $\mu_t$ varies from $1.09$ to $1.36$. }
\end{figure}

\begin{figure}
\begin{center} 
\includegraphics[height=10cm]{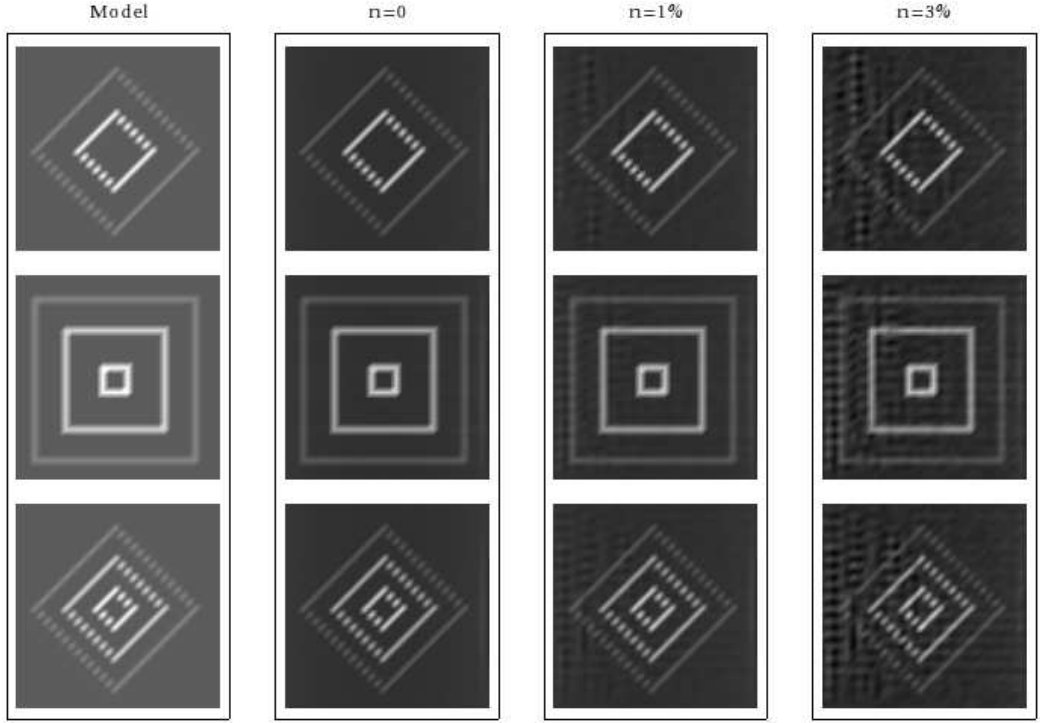}
\end{center}
\caption{\label{mu_a_h_0.04}(Color online) Image reconstruction for the absorption
  coefficient $\mu_a$ for a homogeneously scattering sample with
  $\bar\mu_{s}L_z=1.6$ and $\bar\mu_{a}=0.1\bar\mu_s$, and for various
  noise levels $n$. The rows show the slices $x=6h$, $13h$ and $20h$,
  where the absorbing inhomogeneities are placed. The contrast in
  $\mu_a$ varies from $2$ to $5$. }
\end{figure}

First, we considered the case of purely absorbing inhomogeneities,
spatially modulated as shown in Fig.~\ref{mu_a_h_0.04} in the column
marked ``Model''. The scattering coefficient is constant throughout
the sample and equal to the background value. The absorption
coefficient for the inhomogeneities in the slice $x=6h$ is set to
$\mu_a=2\bar\mu_a$ and $ \mu_a=5\bar\mu_a$, for the outer and inner
square, respectively.  In slices $x=13h$ and $x=20h$ there are more
absorbing inhomogeneities, of absorbing coefficient
$\mu_a=2\bar\mu_a,4\bar\mu_a,5\bar\mu_a$, going from the outmost to
the innermost square. Thus, the contrast of $\mu_t$ (the ratio of
$\mu_t$ in the target to the background value) varies from $1.09$ for
the outmost squares in each slice to $1.36$ for the innermost square.
The results of image reconstruction for the total attenuation
coefficient $\mu_t$ and absorption coefficient $\mu_a$ for various
noise levels $n$ are presented in Figs.~\ref{mu_t_h_0.04} and
\ref{mu_a_h_0.04}. Only the slices containing inhomogeneities are
shown. The other slices present no features, which means that no
cross-talk exists between various slices, as it was also demonstrated
previously \cite{SSOT1}.  It can be seen that the spatial resolution
of images depends on the noise level and can be as good as one
discretization step, $h$.  Note that image reconstruction is in very
good quantitative agreement with the model (all panels in each figure
are plotted using the same color scale) and stable in the presence of
noise.
Note also
that the two-angle measurement scheme considered here enables better
image reconstruction of the attenuation coefficient than the
single-angle scheme considered previously \cite{SSOT1}.

\begin{figure}
\begin{center}
\includegraphics[height=10cm]{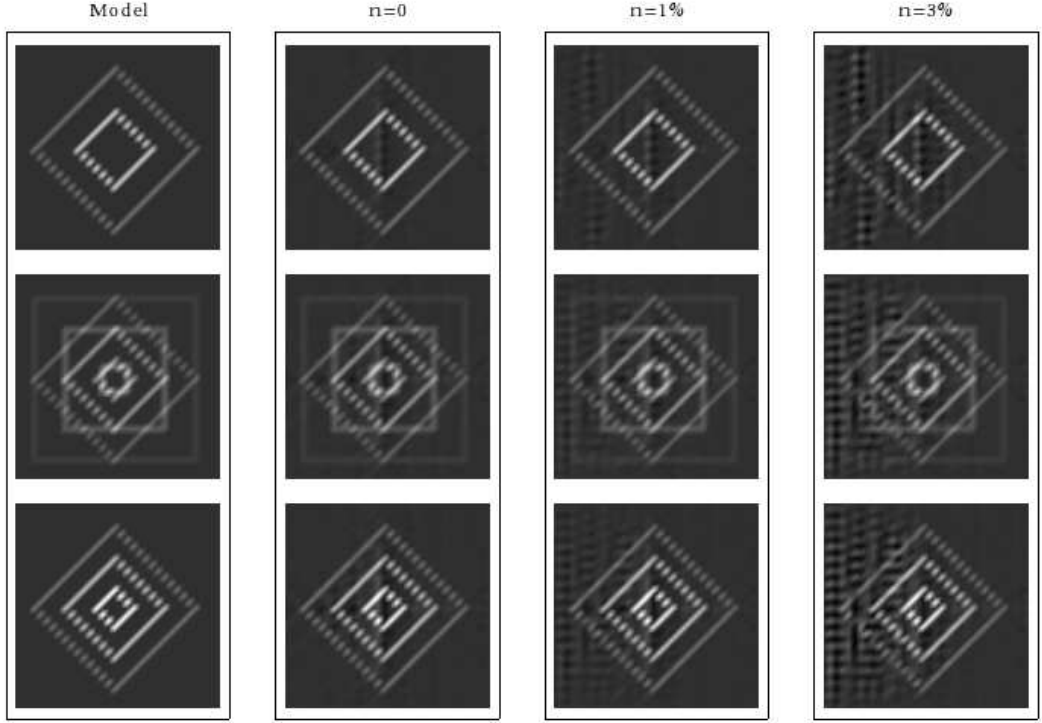}
\end{center}
\caption{\label{mu_t_low_0.04}(Color online) Image reconstruction for the
  total attenuation coefficient $\mu_t$ for an inhomogeneously
  scattering and absorbing sample and for various noise levels
  $n$. The rows show the slices $x=6h$, $13h$ and $20h$, where the
  inhomogeneities are placed. The background scattering and absorption
  coefficients are set such that $\bar\mu_{s}L_z=1.6$ and
  $\bar\mu_{a}=0.1\bar\mu_s$, the contrast in $\mu_s$ varies from
  $1.33$ to $2$, the contrast in $\mu_a$ varies from $2$ to $5$, and
  the contrast in $\mu_t$ varies from $1.09$ to $2.27$.}
\end{figure}

\begin{figure}
\begin{center}
\includegraphics[height=10cm]{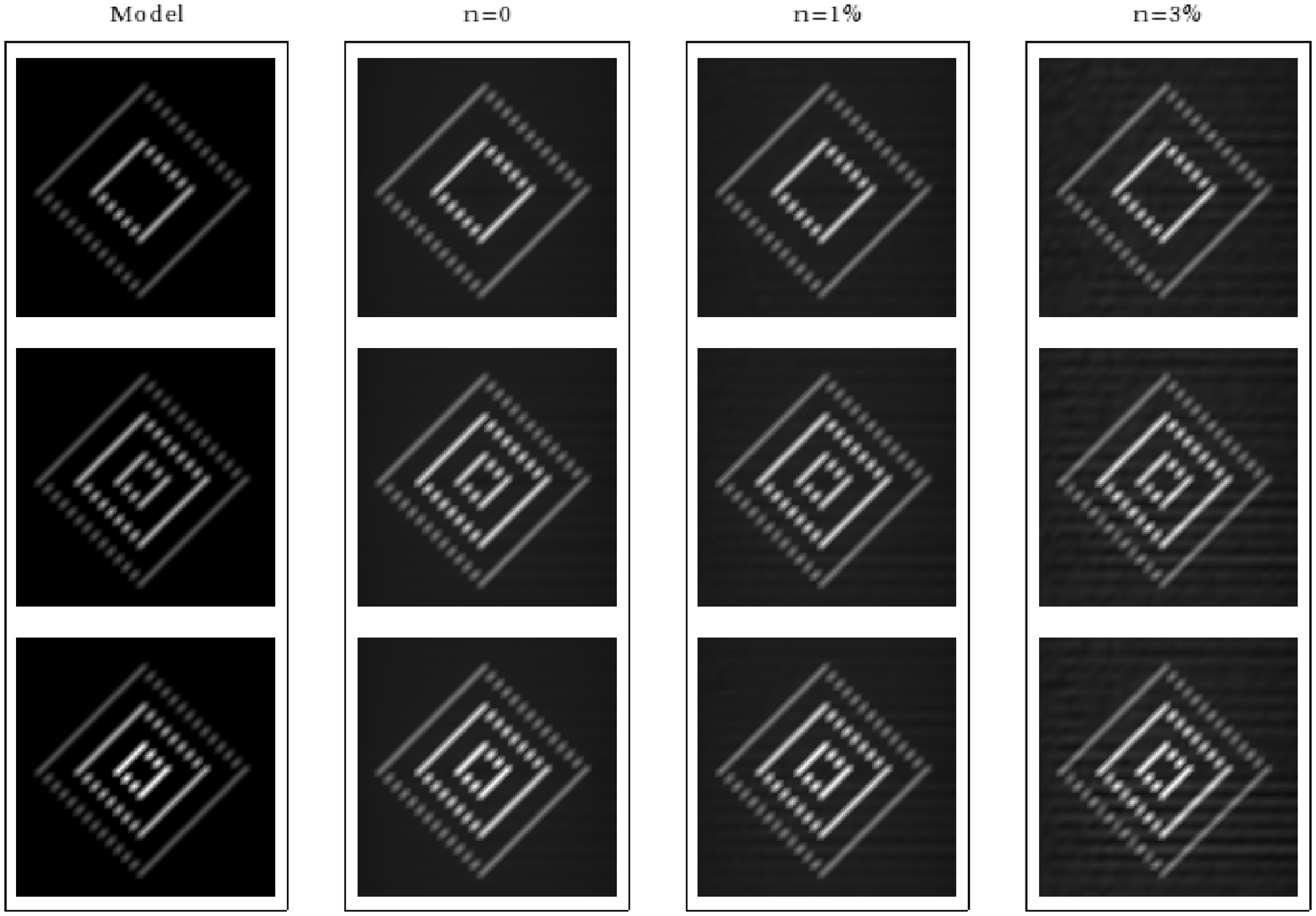}
\end{center}
\caption{\label{mu_s_low_0.04}(Color online)
  Image reconstruction for the scattering coefficient $\mu_s$ for an
  inhomogeneously scattering and absorbing sample and for various
  noise levels $n$. The rows show the slices $x=6h$, $13h$ and $20h$,
  where the inhomogeneities are placed.  The background scattering and
  absorption coefficients are set such that $\bar\mu_{s}L_z=1.6$ and
  $\bar\mu_{a}=0.1\bar\mu_s$, the contrast in $\mu_s$ varies from $1.33$
  to $2$, and the contrast in $\mu_a$ varies from $2$ to $5$.}
\end{figure}

\begin{figure}
\begin{center}
\includegraphics[height=10cm]{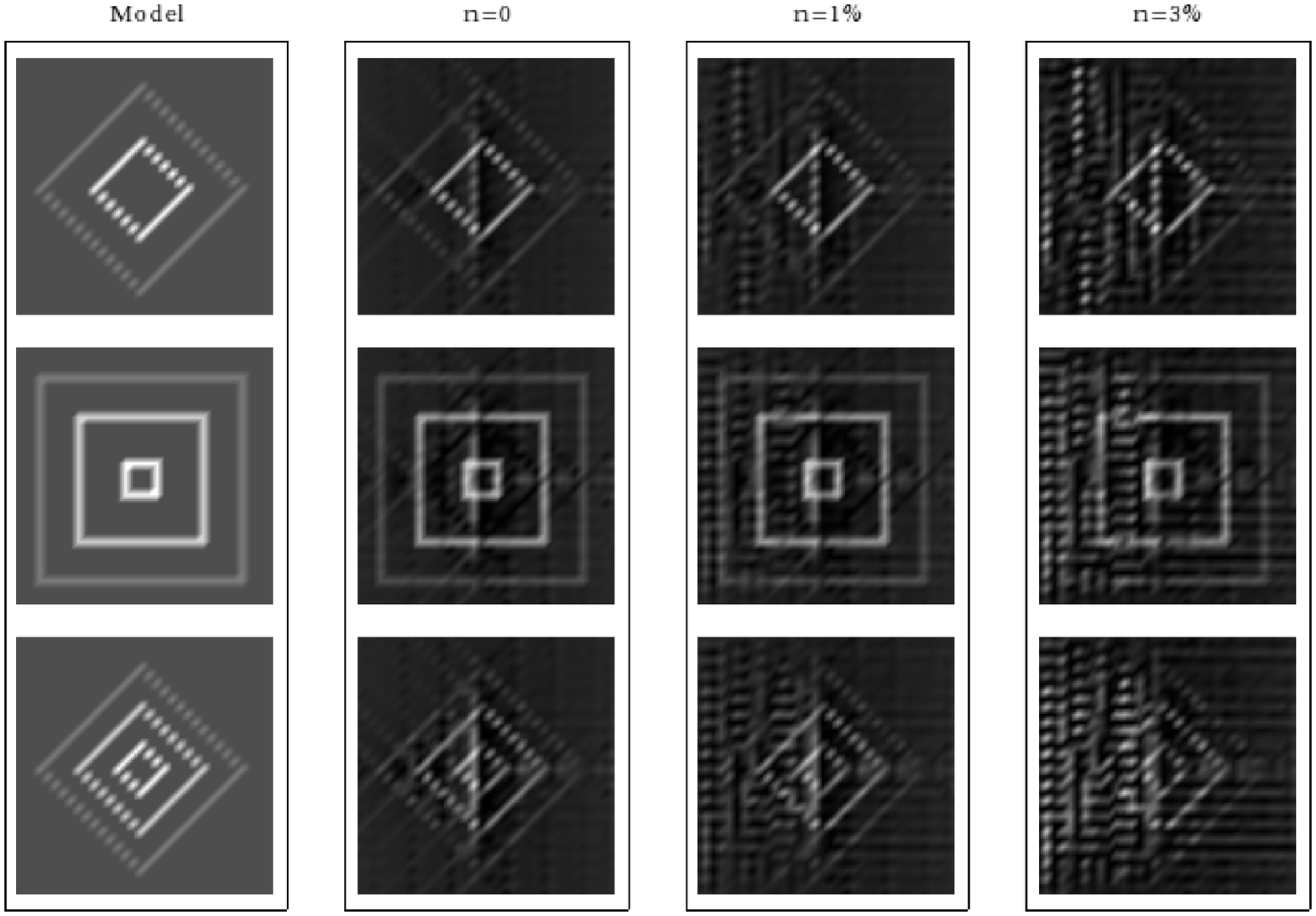}
\end{center}
\caption{\label{mu_a_low_0.04}(Color online)
  image reconstruction for the absorption coefficient $\mu_a$ for an
  inhomogeneously scattering and absorbing sample and for various
  noise levels $n$. The rows show the slices $x=6h$, $13h$ and $20h$,
  where the inhomogeneities are placed.  The background scattering and
  absorption coefficients are set such that $\bar\mu_{s}L_z=1.6$ and
  $\bar\mu_{a}=0.1\bar\mu_s$, the contrast in $\mu_s$ varies from $1.33$
  to $2$, and the contrast in $\mu_a$ varies from $2$ to $5$.}
\end{figure}

Consider now the case when scattering inhomogeneities are also present
in the system and are spatially modulated as shown in
Fig.~\ref{mu_s_low_0.04} in the column marked ``Model'', the
absorption being modulated as described above. The scattering
coefficients for the inhomogeneities in the slice $x=6h$ is set to
$\mu_s=1.33\bar\mu_s$ and $ \mu_s=1.66\bar\mu_s$, for the outer and
inner square, respectively, and in this slice the the absorbing and
scattering inhomogeneities overlap with each other.  In the slice
$x=13h$, there are more scattering inhomogeneities as compared to the
slice $x=6h$, the scattering coefficient is $\mu_s=1.33\bar\mu_s,
1.66\bar\mu_s, 1.66\bar\mu_s$, going from the outermost to the
innermost inhomogeneity, and the absorbing and scattering
inhomogeneities do not overlap. In the slice $x=20h$, the absorbing
and scattering inhomogeneities overlap, the scattering coefficient is
modulated the same as in slice $x=13h$ except that its value for the
innermost inhomogeneity is larger, $\mu_s=2\bar\mu_s$. For this
sample, the contrast of $\mu_t$ varied from $1.09$ for the utmost
squares in slice $x=13$ to $2.27$ for the innermost square in slice
$x=20$.  Imagine reconstruction in this case is presented in
Figs.~\ref{mu_t_low_0.04}-\ref{mu_a_low_0.04}. Very good image quality
is obtained for both the total attenuation coefficient and scattering
coefficient, image reconstruction for scattering being less influenced
by the noise in the data function. 
On the other hand, image quality for absorption is notably lower.
However, most of the relevant features are legible.
 
\begin{figure}
\begin{center}
\includegraphics[height=10cm]{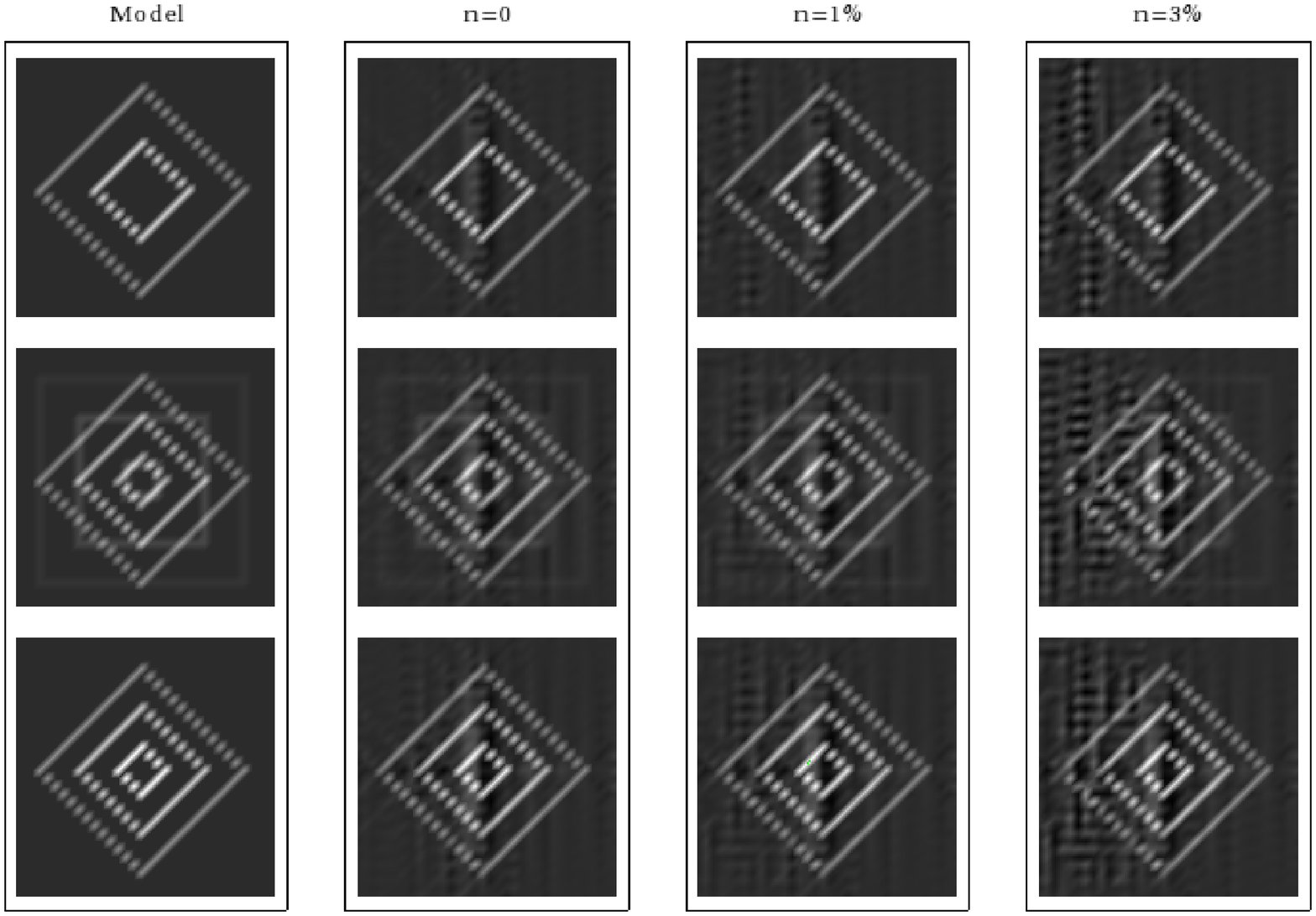}
\end{center}
\caption{\label{mut_0.04}(Color online)
  Image reconstruction for the total attenuation coefficient $\mu_t$ for an
  inhomogeneously scattering and absorbing sample and for various
  noise levels $n$. The rows show the slices $x=6h$, $13h$ and $20h$,
  where the inhomogeneities are placed.  The background scattering and
  absorption coefficients are set such that $\bar\mu_{s}L_z=1.6$ and
  $\bar\mu_{a}=0.1\bar\mu_s$, the contrast in $\mu_s$ varies from
  $2$ to $3$, the contrast in $\mu_a$ varies from $2$ to $5$, and the
  contrast in $\mu_t$ varies from $1.09$ to $3.18$.}
\end{figure}

\begin{figure}
\begin{center}
\includegraphics[height=10cm]{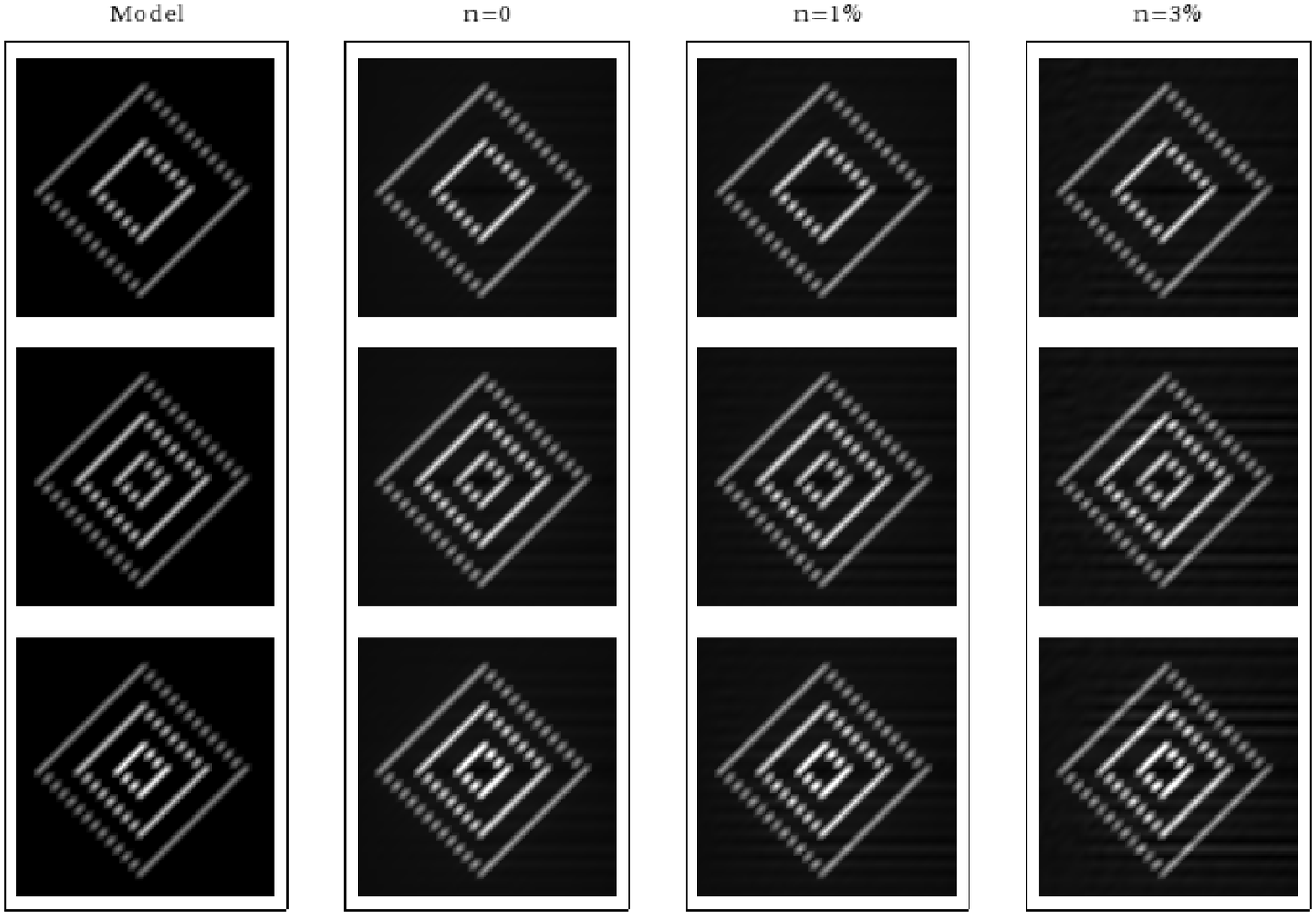}
\end{center}
\caption{\label{mus_0.04}(Color online)
  Image reconstruction for the scattering coefficient $\mu_s$ for an
  inhomogeneously scattering and absorbing sample and for various
  noise levels $n$. The rows show the slices $x=6h$, $13h$ and $20h$,
  where the inhomogeneities are placed.  The background scattering and
  absorption coefficients are set such that $\bar\mu_{s}L_z=1.6$ and
  $\bar\mu_{a}=0.1\bar\mu_s$, the contrast in $\mu_s$ varies from $2$ to
  $3$, and the contrast in $\mu_a$ varies from $2$ to $5$. }
\end{figure}

\begin{figure}
\begin{center}
\includegraphics[height=10cm]{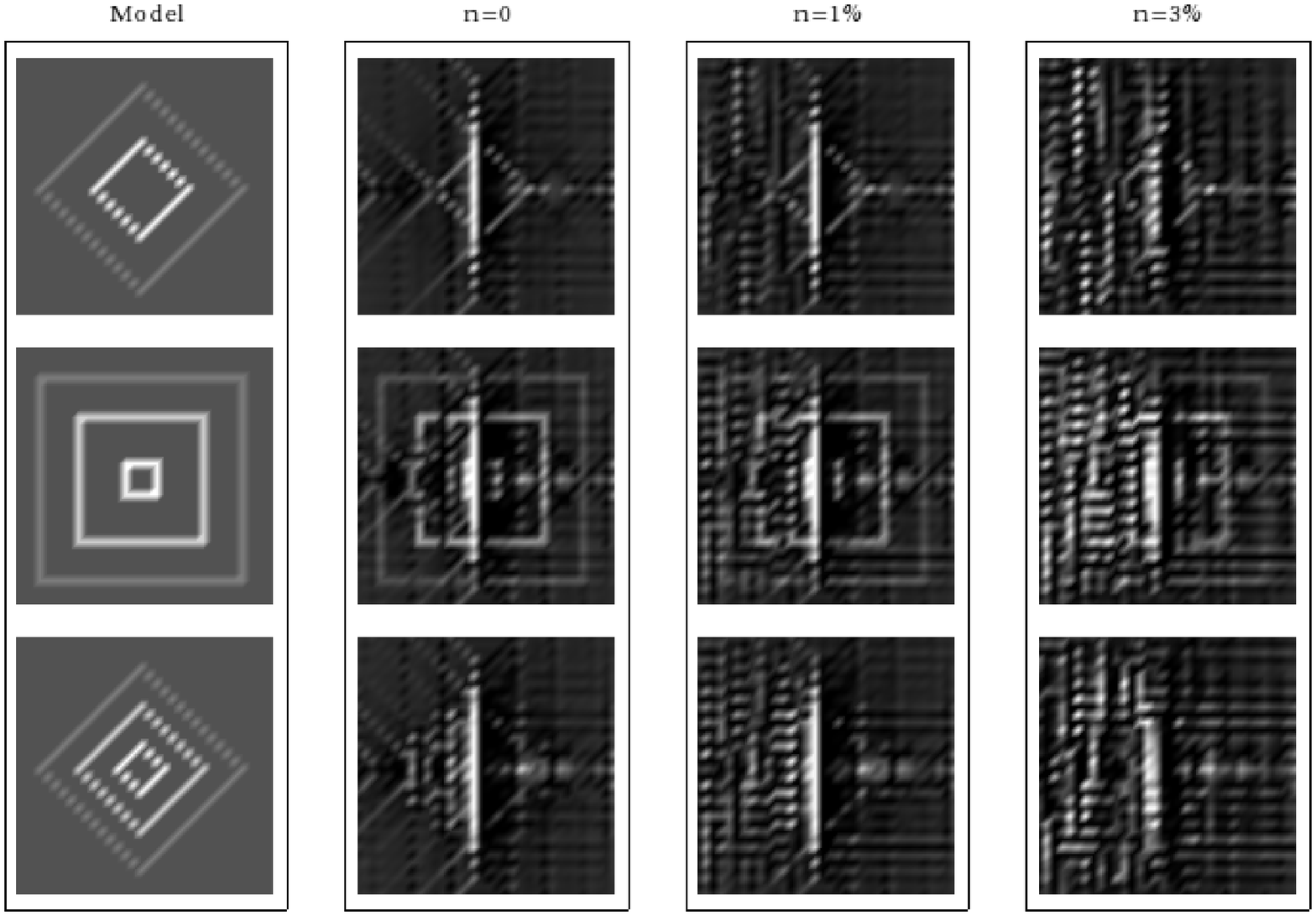}
\end{center}
\caption{\label{mua_0.04}(Color online) Image reconstruction for the
  absorption coefficient $\mu_a$ for an inhomogeneously scattering and
  absorbing sample and for various noise levels $n$. The rows show the
  slices $x=6h$, $13h$ and $20h$, where the inhomogeneities are
  placed.  The background scattering and absorption coefficients are
  set such that $\bar\mu_{s}L_z=1.6$ and $\bar\mu_{a}=0.1\bar\mu_s$,
  the contrast in $\mu_s$ varies from $2$ to $3$, and the contrast in
  $\mu_a$ varies from $2$ to $5$. }
\end{figure}

Further, consider stronger scattering inhomogeneities and perform
image reconstruction for a sample in which the scattering coefficient
of the inhomogeneities is spatially modulated the same as in the
previous case, but it is increased by a factor of $1.5$, the absorbing
inhomogeneities having the same characteristics.  Thus, in this case,
the contrast in the scattering coefficient varies from $2$ for the
outmost inhomogeneity in each slice to $3$ for the innermost
inhomogeneity in slice $x=20$. The contrast in $\mu_t$ varies from
$1.09$ to $3.18$. The results are presented in
Figs.~\ref{mut_0.04}-\ref{mua_0.04}. Very good image reconstruction is
obtained for both the total attenuation and scattering coefficients,
but imagine reconstruction for absorption is very poor.

\begin{figure}
\begin{center}
\includegraphics[height=10cm]{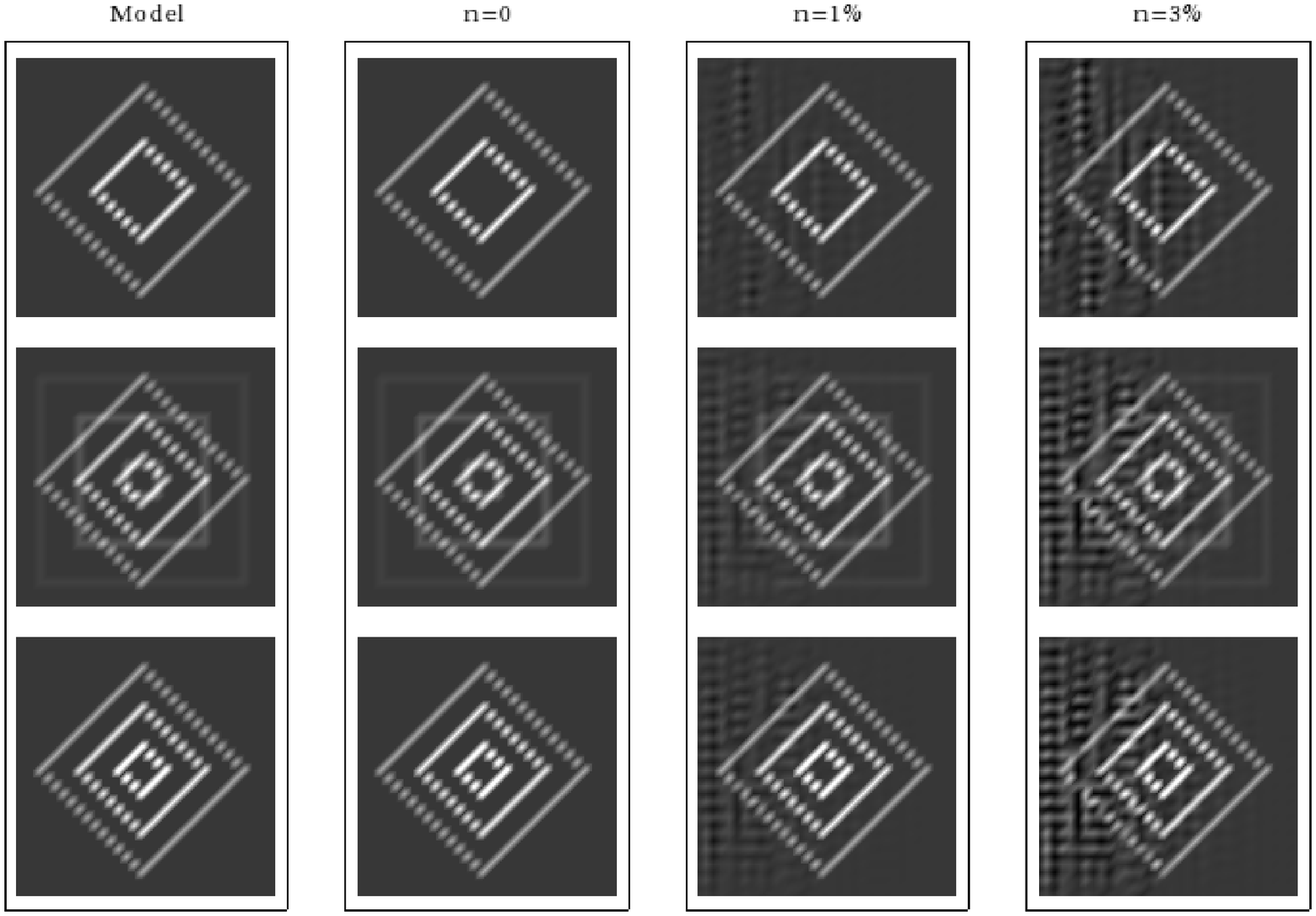}
\end{center}
\caption{\label{mut_0.04_ik}(Color online)
  Image reconstruction for the total attenuation coefficient  $\mu_t$ for various noise
  levels $n$, for a data function
  corresponding only to single-scattered light and calculated
  according with (\ref{SIM_SSOT_eq1}) (inverse crime). All the sample parameters are
  as for Fig.~\ref{mut_0.04}.}
\end{figure}

\begin{figure}
\begin{center}
\includegraphics[height=10cm]{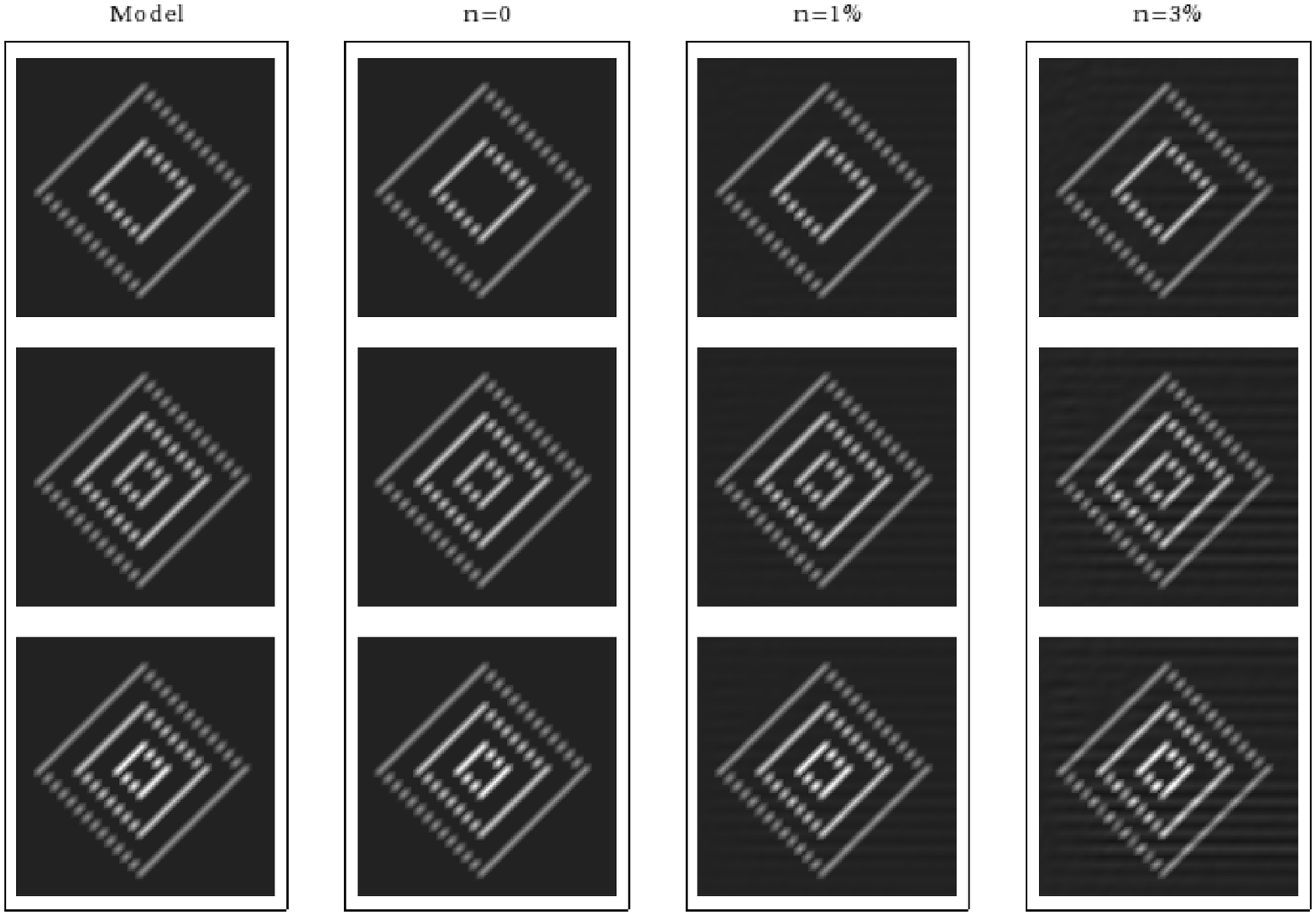}
\end{center}
\caption{\label{mus_0.04_ik}(Color online)
  Image reconstruction for the scattering coefficient $\mu_s$ for
  various noise levels $n$, for a data function corresponding only to
  single-scattered light and calculated according with
  (\ref{SIM_SSOT_eq1}) (inverse crime). All the sample parameters are
  as for Fig.~\ref{mus_0.04}.}
\end{figure}

\begin{figure}
\begin{center}
\includegraphics[height=10cm]{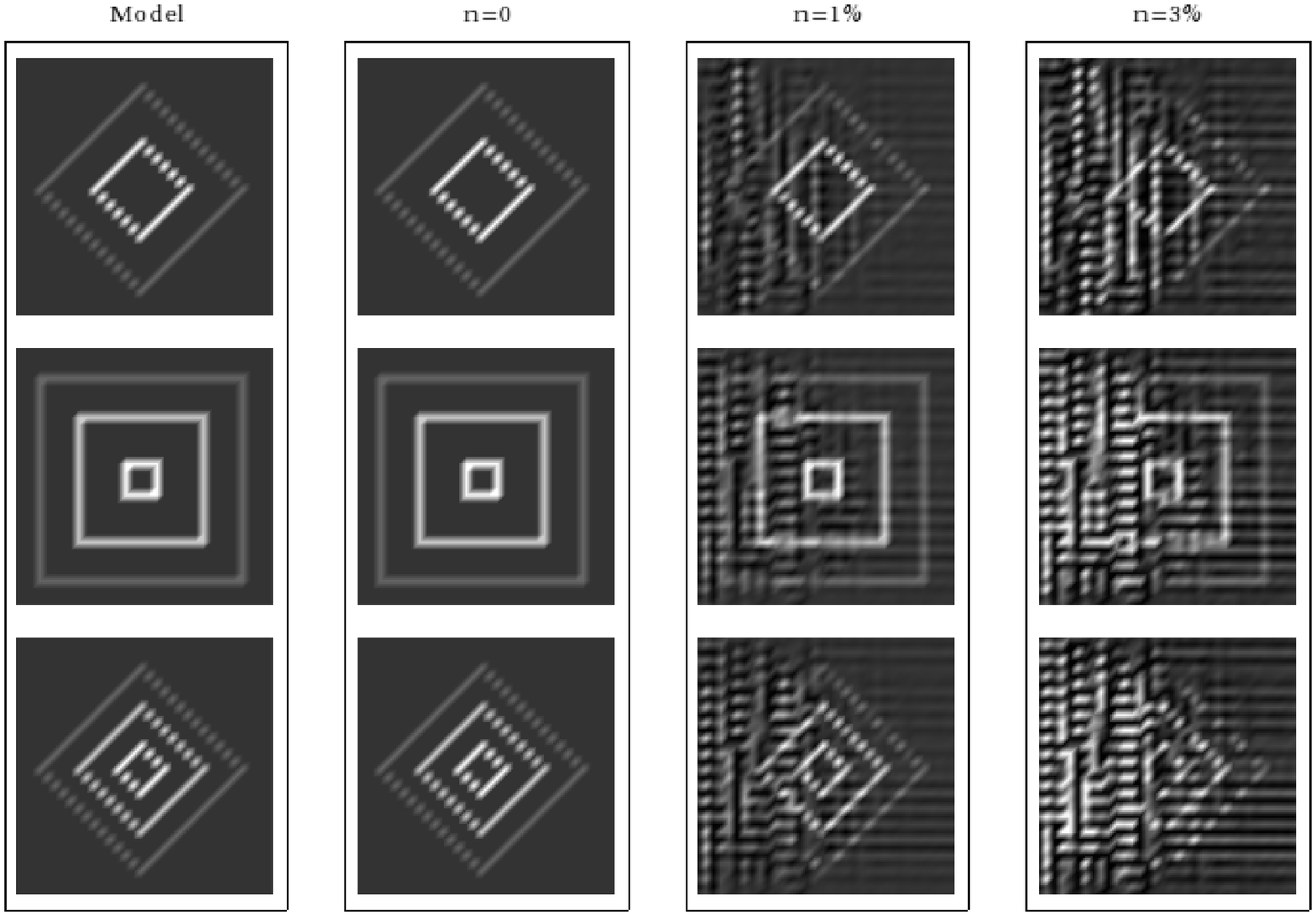}
\end{center}
\caption{\label{mua_0.04_ik}(Color online)
  Image reconstruction for the absorption coefficient $\mu_a$ for
  various noise levels $n$, for a data function corresponding only to
  single-scattered light and calculated according with
  (\ref{SIM_SSOT_eq1}) (inverse crime). All the sample parameters are
  as for Fig.~\ref{mua_0.04}.}
\end{figure}

The reconstructed image quality is determined by two factors, the
amount of scattering in the system and the noise in the data. In
particular, for stronger scattering, the single-scattering
approximation we employ may be inaccurate, leading to poor image
reconstruction.  In order to separate the influence of these factors
on the image quality, we perform image reconstruction based on a data
function corresponding only to single-scattered light, obtained using
the so-called inverse crime. This consists of generating data using
the same model that the inverse solver is based on.  Specifically,
instead of solving RTE numerically and using the solution to calculate
the data function according to the definition (\ref{data_def}), the
data function is calculated from (\ref{SIM_SSOT_eq1}), derived within
the single-scattering approximation of RTE, by replacing the
extinction and scattering coefficients by those of the model. In this
case, the influence of the amount of scattering in the sample on the
image quality is eliminated, the only influence coming from the noise
in the data.  Image reconstruction for the same sample that was
analyzed in Figs.~\ref{mut_0.04}-\ref{mua_0.04} is presented in
Figs.~\ref{mut_0.04_ik}-\ref{mua_0.04_ik}.  
 By comparing these two sets of results, one can see that the imagine
quality for the attenuation and scattering coefficients is almost the
same in both cases. Therefore, it can be concluded that the
single-scattering approximation works very well for the scattering
strength considered and that, in this scattering regime, the most
influence on the image quality comes from the noise in the data.
Moreover, Figs.~\ref{mut_0.04_ik}-\ref{mua_0.04_ik} show that the
reconstructed coefficients experience various levels of influence from
the noise in the data. While the reconstructed attenuation and
scattering coefficients are very little influenced by the noise, the
absorption coefficient experiences a much stronger influence, imagine
quality being very poor even for a noise level of $3\%$ in the
measured intensity.

To explain the various levels of influence of the noise in the data on
the image quality, we perform a rough error propagation analysis.
Assume that the scattered intensity $I_s$ is determined with an error
$\delta I_s$.  From Eq.~(\ref{mu_SVD}) and the definition
(\ref{data_def}) of the data function, it follows that the noise in
the data results in a noise $\delta \mu_{tn}$ in the total attenuation
coefficient given by 
\begin{equation}
\label{delta_mut}
\delta \mu_{tn}= \sqrt 2\,\frac{\delta
I_s}{I_s}\,\left(\sum_m\frac{|f_{mn}|^2}{\sigma_m^{2}}\Theta(\sigma_m^2
- \epsilon)\right)^{1/2}.
\end{equation}
Here, we have assumed that the relative error in determining the
scattered intensity is the same for the symmetric measurements used to
reconstruct $\mu_t$, and the factor $\sqrt 2$ results from using the
the difference in the data functions corresponding to these
measurements. On the other hand, the noise $\delta \mu_{sn}$ in the
scattering coefficient $\mu_{sn}$ is estimated from
Eqs.~(\ref{SSOT_2D_d}) and (\ref{delta_mut}) to be
\begin{equation}
\label{delta_mus}
\frac{\delta \mu_{sn}}{\mu_{sn}}= \frac{\delta
  I_s}{I_s}\left(1+2\sum_i(L_{\nu i}^{(1)})^2\sum_m\frac{|g_{mi}|^2}{\sigma_m^{2}}\Theta(\sigma_m^2 -
\epsilon)\right)^{1/2}.
\end{equation}
Here $n$ labels the cell where the detected rays corresponding to the
source-detection realization $\nu$ are single scattered. Finally, the
absorption coefficient is determined as the difference between the
total attenuation coefficient and the scattering coefficient with the
error $\delta \mu_{an}$ that verifies
\begin{equation}
\frac{\delta \mu_{an}}{\mu_{an}}
  =\frac{\mu_{sn}}{\mu_{an}}\left[\left(\frac{\delta
  \mu_{sn}}{\mu_{sn}}\right)^2+\left(\frac{\mu_{tn}}{\mu_{sn}}\right)^2\,\left(\frac{\delta
  \mu_{tn}}{\mu_{tn}}\right)^2\right]^{1/2}>\frac{\mu_{sn}}{\mu_{an}}\frac{\delta
  I_s}{I_s}.
\end{equation}
This expresses the fact that for samples where the absorption is
stronger than the scattering, very large noise to signal levels in the
reconstructed absorption coefficient result, even for low noise levels
in the data.
This result completely explains the image reconstruction for the
absorption coefficient presented above. The difference between
Figs.~\ref{mu_a_low_0.04} and \ref{mua_0.04} is that the maximum value
of the ratio $\mu_s/\mu_a$ is increased from $4$ (for
Fig.~\ref{mu_a_low_0.04}) to $6$ (for Fig.~\ref{mua_0.04}), leading to
pronounced noise in the reconstructed image. Also, although stronger
scattering inhomogeneities are present in slice $x=13$ (second row) in
Fig.~\ref{mua_0.04}, better image quality is obtained compared with
the slice $x=6$ (first row), since in this case the absorbing and
scattering inhomogeneities do not overlap and $\mu_s/\mu_a$ is
smaller.  The physical interpretation of this result is that the SSOT
data function carries more signature of the scattering coefficient
than of the absorption. In other words, in SSOT, the scattering
coefficient is privileged as compared with the absorption coefficient.
This fact originates from the RTE equation we employ, where the
scattering coefficient has a stronger contribution than the absorption
coefficient, and is also expressed by Eqs.~(\ref{data_def}) and
(\ref{SIM_SSOT_eq1}), showing that the scattered intensity decays
exponentially with the absorption coefficient, but has a stronger
dependence on the scattering coefficient. In this case, besides the
attenuation of the specific intensity as a result of absorption and
scattering of photons from a given mode into other modes, there is
also amplification of a given mode due to scattering of photons from
other modes into that mode. We note that this dependence of the
measured intensity of the scattering and absorption coefficients of
the sample is different from the case of diffuse optical
tomography. In the case of diffusive light propagation, the scattering
of photons into and out various directional modes does not affect the
light intensity, the diffusion equation has similar contributions from
the scattering and absorption coefficients, and the light intensity
emerging from the sample decays exponentially with both $\mu_s$ and
$\mu_a$.

\begin{figure}
\begin{center}
\includegraphics[height=10cm]{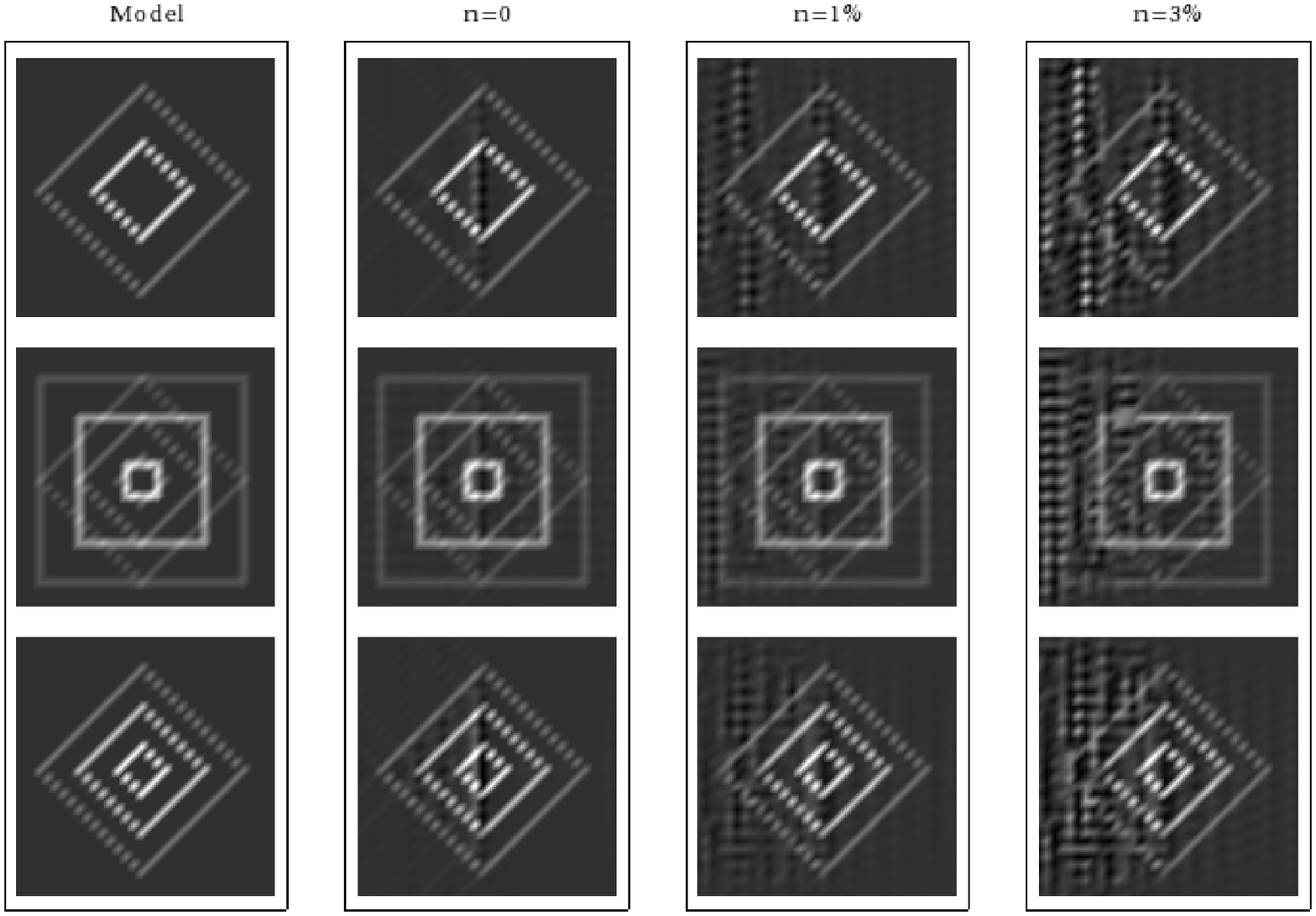}
\end{center}
\caption{\label{mut_0.04_ha}(Color online)
   Image reconstruction for the total attenuation coefficient $\mu_t$ for an
  inhomogeneously scattering and absorbing sample and for various
  noise levels $n$. The rows show the slices $x=6h$, $13h$ and $20h$,
  where the inhomogeneities are placed.  The background scattering and
  absorption coefficients are set such that $\bar\mu_{s}L_z=1.6$ and
  $\bar\mu_{a}=\bar\mu_s$, the contrast in $\mu_s$ varies from
  $2$ to $3$, the contrast in $\mu_a$ varies from $2$ to $5$, and the
  contrast in $\mu_t$ varies from $2$ to $4$.}
\end{figure}

\begin{figure}
\begin{center}
\includegraphics[height=10cm]{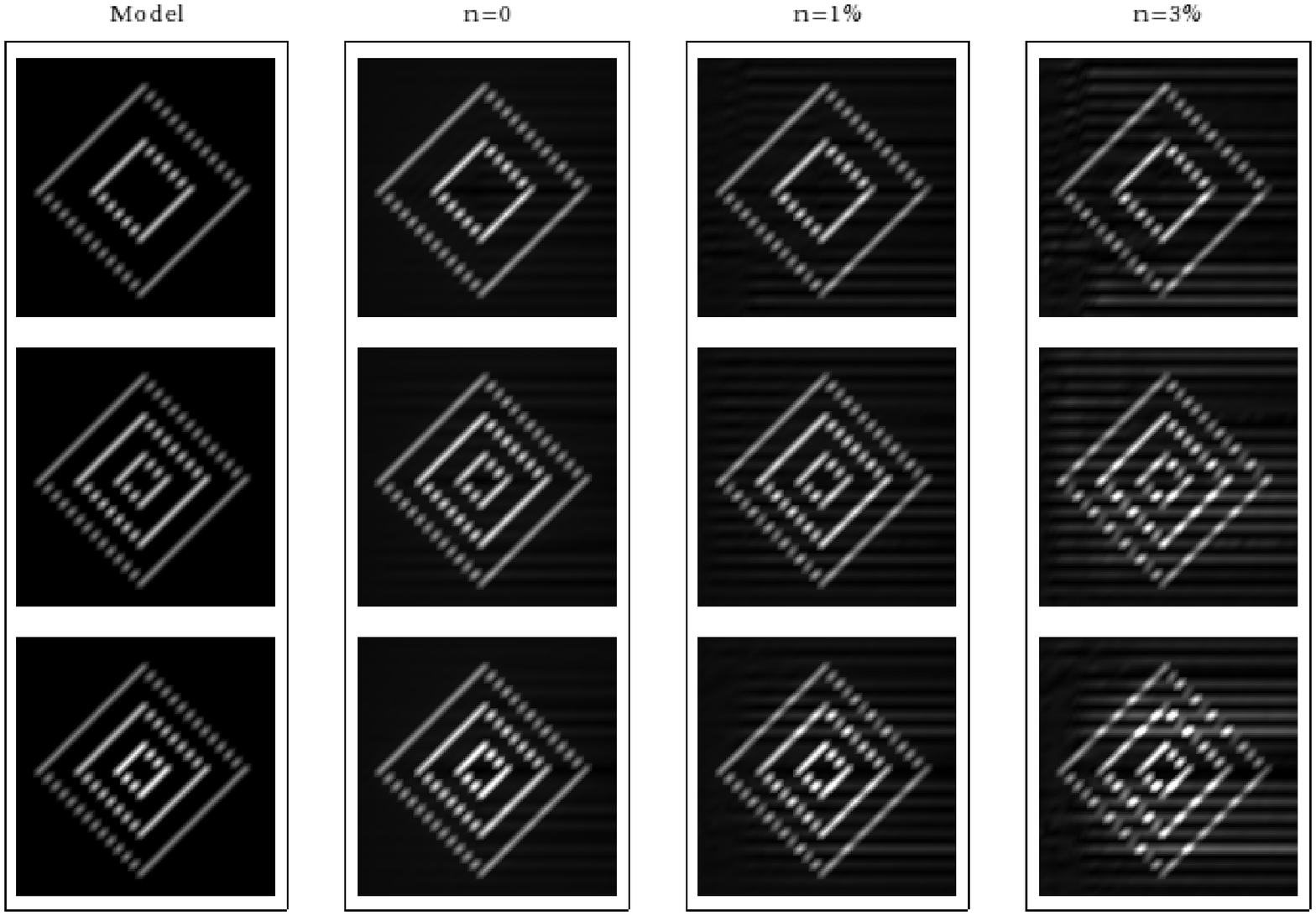}
\end{center}
\caption{\label{mus_0.04_ha}(Color online)
 Image reconstruction for the scattering coefficient $\mu_s$ for an
  inhomogeneously scattering and absorbing sample and for various
  noise levels $n$. The rows show the slices $x=6h$, $13h$ and $20h$,
  where the inhomogeneities are placed.  The background scattering and
  absorption coefficients are set such that $\bar\mu_{s}L_z=1.6$ and
  $\bar\mu_{a}=\bar\mu_s$, the contrast in $\mu_s$ varies from $2$ to
  $3$, and the contrast in $\mu_a$ varies from $2$ to $5$. }
\end{figure}

\begin{figure}
\begin{center}
\includegraphics[height=10cm]{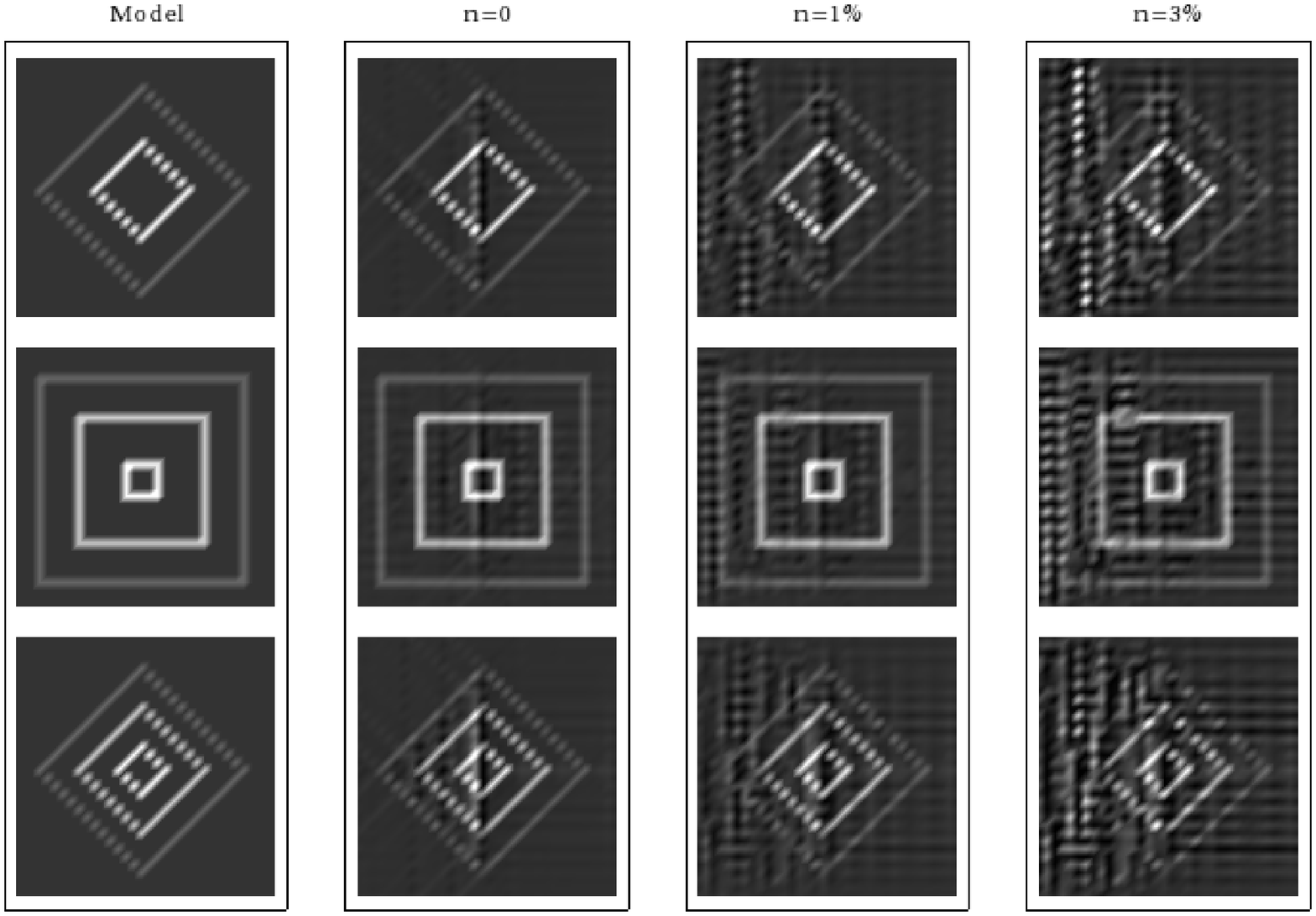}
\end{center}
\caption{\label{mua_0.04_ha}(Color online)
 Image reconstruction for the
  absorption coefficient $\mu_a$ for an inhomogeneously scattering and
  absorbing sample and for various noise levels $n$. The rows show the
  slices $x=6h$, $13h$ and $20h$, where the inhomogeneities are
  placed.  The background scattering and absorption coefficients are
  set such that $\bar\mu_{s}L_z=1.6$ and $\bar\mu_{a}=\bar\mu_s$,
  the contrast in $\mu_s$ varies from $2$ to $3$, and the contrast in
  $\mu_a$ varies from $2$ to $5$. }
\end{figure}

To verify the conclusion presented above, we perform image
reconstruction for a stronger absorbing sample.
Figs.~\ref{mut_0.04_ha}-\ref{mua_0.04_ha} present image reconstruction
for a sample where the absorption and scattering are spatially
modulated as for Figs.~\ref{mut_0.04}-\ref{mua_0.04}, but the
absorption coefficient is increased by a factor of $10$, such that it
becomes comparable to the scattering coefficient. Indeed, the image
reconstruction for the absorption coefficient presented in
Fig.~\ref{mua_0.04_ha} is markedly better than in Fig.~\ref{mua_0.04}
and comparable to that for the scattering  coefficient
presented in Fig.~\ref{mus_0.04_ha}.


\begin{figure}
\begin{center}
\includegraphics[height=10cm]{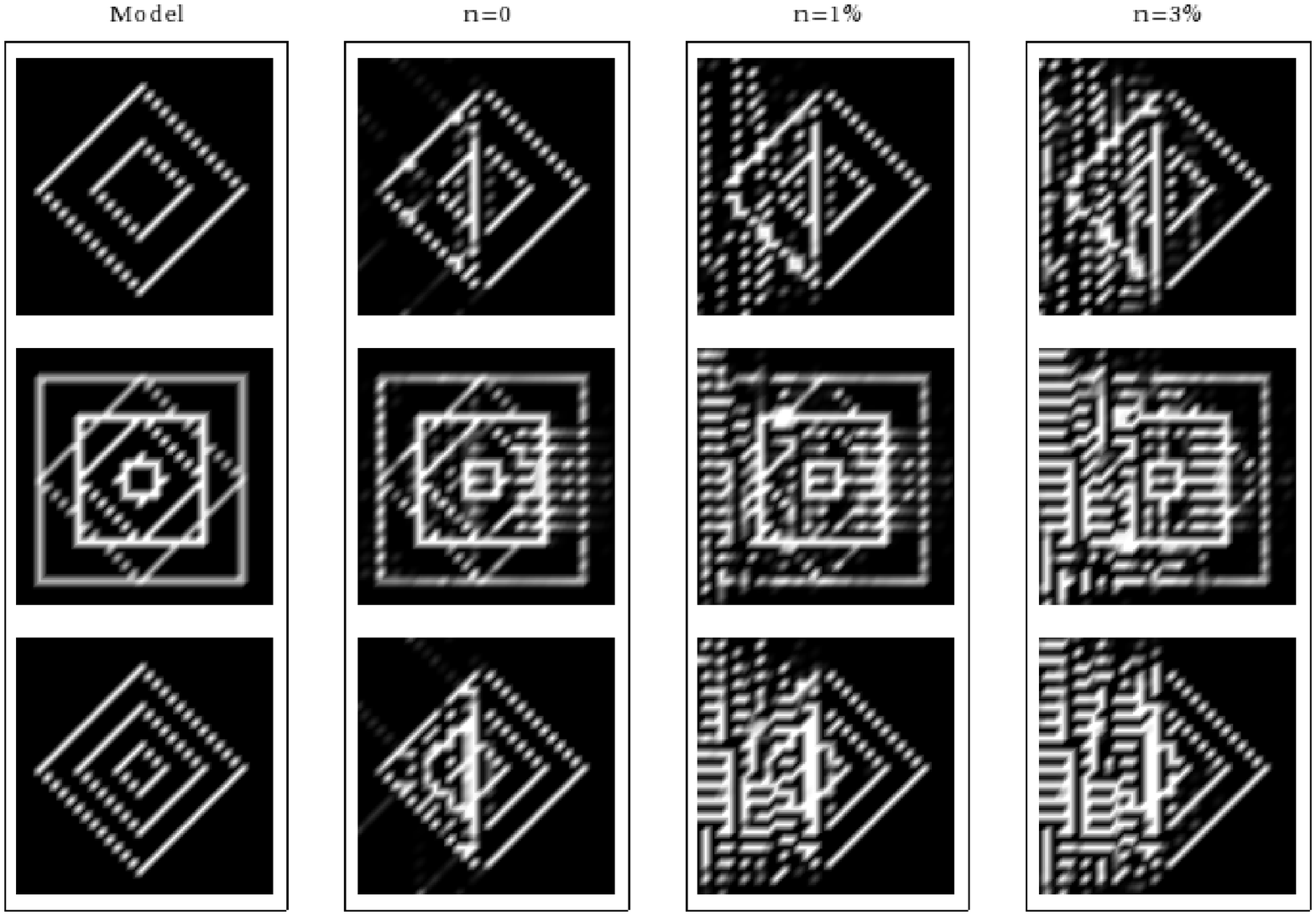}
\end{center}
\caption{\label{mut_0.08_ha}(Color online)
   Image reconstruction for the total attenuation coefficient $\mu_t$ for an
  inhomogeneously scattering and absorbing sample and for various
  noise levels $n$. The rows show the slices $x=6h$, $13h$ and $20h$,
  where the inhomogeneities are placed.  The background scattering and
  absorption coefficients are set such that $\bar\mu_{s}L_z=3.2$ and
  $\bar\mu_{a}=\bar\mu_s$, the contrast in $\mu_s$ varies from
  $2$ to $3$, the contrast in $\mu_a$ varies from $2$ to $5$, and the
  contrast in $\mu_t$ varies from $2$ to $4$.}
\end{figure}

\begin{figure}
\begin{center}
\includegraphics[height=10cm]{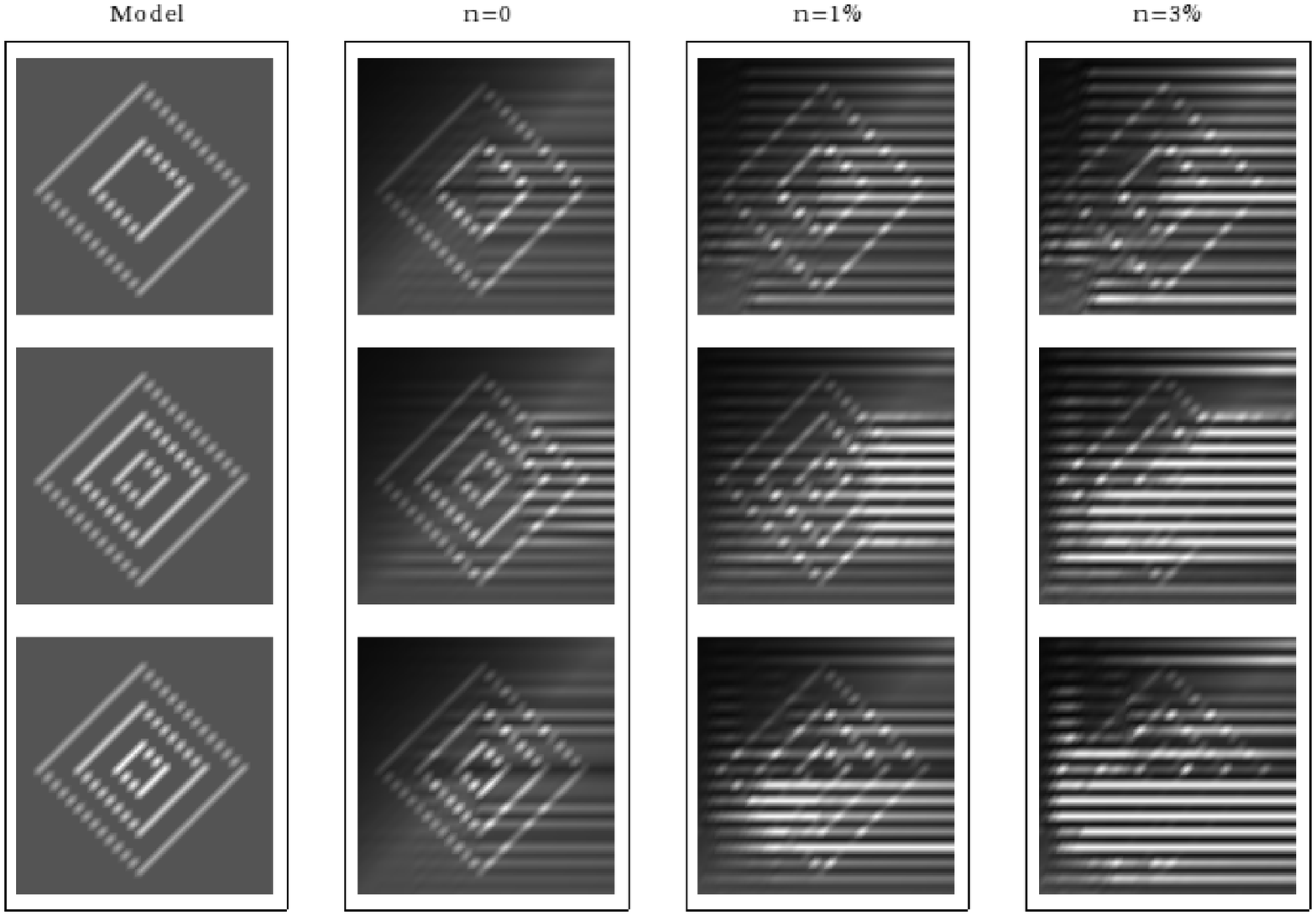}
\end{center}
\caption{\label{mus_0.08_ha}(Color online)
 Image reconstruction for the scattering coefficient $\mu_s$ for an
  inhomogeneously scattering and absorbing sample and for various
  noise levels $n$. The rows show the slices $x=6h$, $13h$ and $20h$,
  where the inhomogeneities are placed.  The background scattering and
  absorption coefficients are set such that $\bar\mu_{s}L_z=3.2$ and
  $\bar\mu_{a}=\bar\mu_s$, the contrast in $\mu_s$ varies from $2$ to
  $3$, and the contrast in $\mu_a$ varies from $2$ to $5$. }
\end{figure}

\begin{figure}
\begin{center}
\includegraphics[height=10cm]{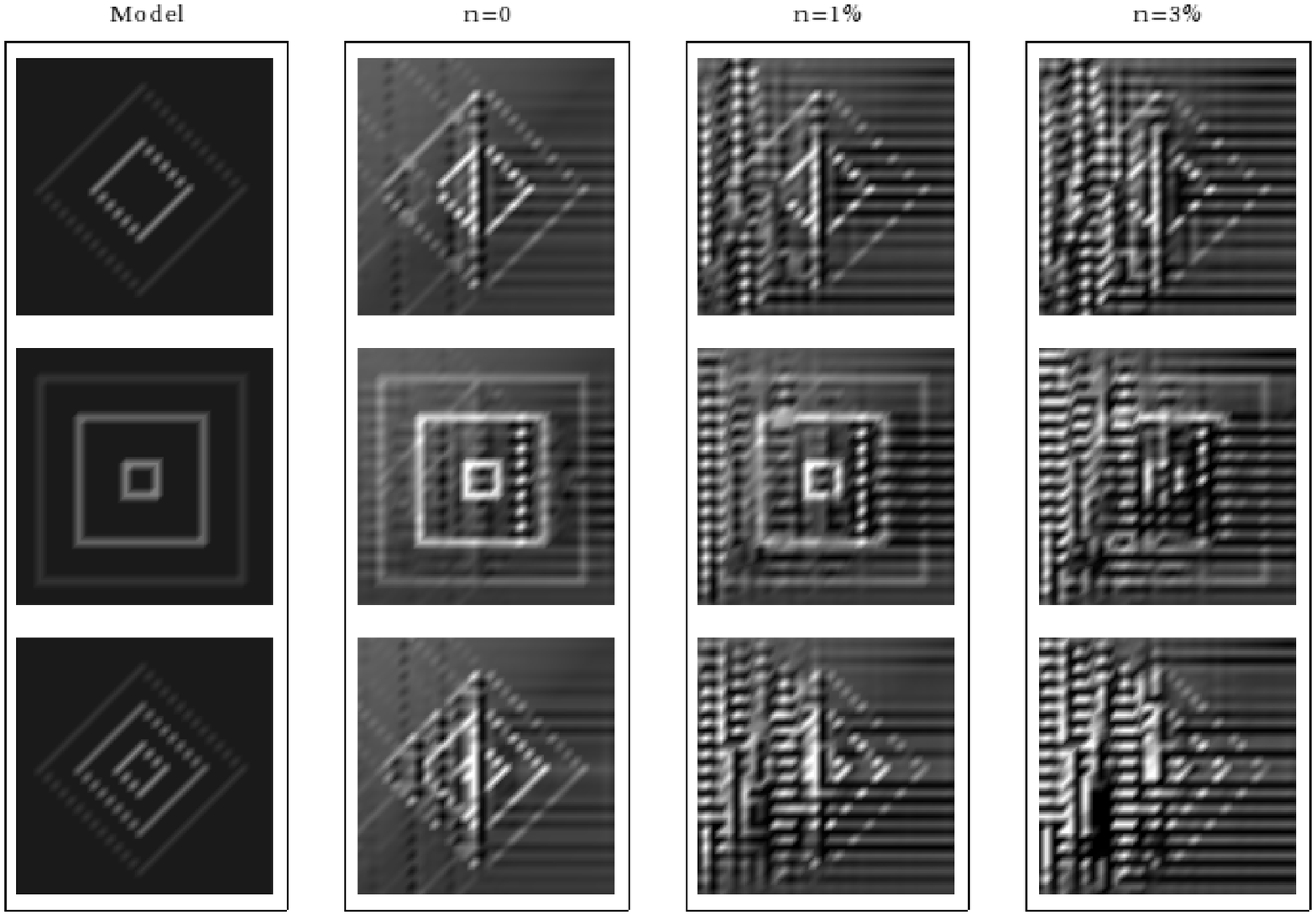}
\end{center}
\caption{\label{mua_0.08_ha}(Color online)
 Image reconstruction for the
  absorption coefficient $\mu_a$ for an inhomogeneously scattering and
  absorbing sample and for various noise levels $n$. The rows show the
  slices $x=6h$, $13h$ and $20h$, where the inhomogeneities are
  placed.  The background scattering and absorption coefficients are
  set such that $\bar\mu_{s}L_z=3.2$ and $\bar\mu_{a}=\bar\mu_s$,
  the contrast in $\mu_s$ varies from $2$ to $3$, and the contrast in
  $\mu_a$ varies from $2$ to $5$. }
\end{figure}

Finally, we perform image reconstruction for a stronger scattering
sample, characterized by an optical depth of the background of
$\bar\mu_sL_z=3.2$ and an additional contrast in the scattering
coefficient of up to $3$. This is a borderline case when scattering is
sufficiently strong so that the single-scattering approximation of
SSOT may be expected to be inaccurate. The results for imagine
reconstruction obtained for the case when the scattering and
absorption have comparable strengths are presented in
Figs.~\ref{mut_0.08_ha}-\ref{mua_0.08_ha}. We obtain that even in this
scattering regime the most relevant features in the reconstructed
scattering and absorption coefficients remain legible.


\section{Conclusions}
\label{sec:conclusions}
We have demonstrated that the SSOT technique enables simultaneous
reconstruction of scattering and absorption properties of mesoscopic
systems. In particular, we have shown that while accurate, qualitative
imagine reconstruction of scattering is always possible, good image
reconstruction for absorption can be realized under the condition that
scattering and absorption have comparable strengths.  These
conclusions have been reached under the assumption that the light
propagating in the mesoscopic systems is just single scattered, but
without making any assumption of measuring just single-scattered
light. We have argued that better image quality for scattering as
compared to absorption is possible since the SSOT data function
carries a stronger signature of scattering than of absorption.

Simultaneous reconstruction of scattering and absorption of mesoscopic
systems can be experimentally implemented by appropriately choosing
the wavelength of the illuminating beam, such that the effects of
absorption and scattering have comparable strengths
\cite{vanGermert_book}. Alternatively, the absorption characteristics
of the sample can be recovered through fluorescent SSOT, which will be
the subject of a future study.

\section*{Acknowledgment}
This work was supported by the National Science Foundation under Grant
No. 0615857

\end{document}